\documentclass[aps,groupedaddress,superscriptaddress,amsmath,amssymb,twocolumn,pra]{revtex4-2}
\usepackage{bm}
\usepackage[pdftex]{graphicx}
\usepackage{xcolor}
\usepackage{color}
\usepackage[normalem]{ulem}
\usepackage{enumerate}
\usepackage{epstopdf}
\usepackage{braket}

\def\Bperp{B_\perp}
\def\Bp{B_\parallel}

\def\kb{{\bf k}}

\def\rb{{\bf r}}
\def\Rb{\mathbf{R}}
\def\Xb{\mathbf{X}}
\def\pb{{\bf p}}

\def\Eb{\mathbf{E}}

\def\omp{\omega_\parallel}
\def\ep{\varepsilon}

\newcommand{\sumprime}[1]{\sum_{#1}{\vphantom{\sum}}^{\!\!\prime}}
\def\n{{\bm\nabla}}

\begin{document}

\title{Many-electron system on helium and the color center spectroscopy}

\author{A. D. Chepelianskii}
\affiliation{LPS, Universit\'e Paris-Saclay, CNRS, UMR 8502, F-91405 Orsay, France}
\author{ D. Konstantinov}
\affiliation{Okinawa Institute of Science and Technology (OIST) Graduate University, Onna, Okinawa 904-0412, Japan}
\author{M. I. Dykman}
\affiliation{Department of Physics and Astronomy, Michigan State University, East Lansing, MI 48824, USA}

\begin{abstract}
Electrons on the helium surface display sharp resonant absorption lines related to the transitions between the subbands of quantized motion transverse to the surface. A magnetic field parallel to the surface strongly affects the absorption spectrum.  We show that the effect comes from admixing the out-of-plane motion to the in-plane quantum dynamics of the strongly correlated  electron liquid or a Wigner crystal. This is similar to the admixing electron transitions in color centers to phonons. The spectrum permits a direct characterization of the many-electron dynamics and also enables testing the theory of color centers in a system with a controllable coupling. 
\end{abstract}

\date{\today}

\maketitle

Electrons above the surface of liquid helium are localized in a one-dimensional potential well, which is formed by the high repulsive barrier at the surface and the image potential. 
The energy levels in the well are quantized.
The electrons occupy the lowest level forming a two-dimensional system \cite{Andrei1997,Monarkha2004}. 
The spectroscopic observation of transitions between the quantized energy levels \cite{Grimes1976a} was a direct proof of the picture of the electron confinement and the overall nature of the potential. Since then much work has been done on the exact positions and the widths of the spectral lines and their dependence on the temperature and the electron density \cite{Stern1978,Lambert1980,RamaKrishna1988,Cheng1994,Collin2002,Degani2005,Konstantinov2009,Dykman2017,Yunusova2019}. 

The electron system on helium is free from static disorder. It is also weakly coupled to the vibrational excitations in helium, ripplons and phonons. The observed spectral lines are narrow, with width as small as  $\sim 2$~MHz for $T=0.3$~K \cite{Collin2002}. The electron transitions are not accompanied by creating ripplons or phonons. In the nomenclature of the solid-state spectroscopy they correspond to zero-phonon  lines. Such lines in the spectra of point defects result from transitions between the defect energy levels with no energy transfer to/from phonons \cite{Stoneham2001}. The physics of point defects and the defect spectroscopy have been in the focus of attention recently in the context of quantum computing and quantum sensing \cite{Awschalom2018}. On their side, electrons on helium themselves have been  also considered as a viable candidate system for a scalable quantum computer \cite{Platzman1999,Lyon2006,Schuster2010,Yang2016a,Byeon2020}. 

One of the major attractive features of electrons on helium is the possibility to study many-electron effects. 
The electron-electron interaction is strong, the ratio of its energy to the electron kinetic energy is $\Gamma = e^2(\pi n_s)^{1/2}/k_BT >30$ for the electron density $n_s\geq 10^7~\mathrm{cm}^{-2}$ and $T\leq 0.3$~K. The electrons form a Wigner crystal \cite{Grimes1979,Fisher1979}  or a classical or quantum nondegenerate liquid with unusual transport properties, cf.  \cite{Dykman1979a,Edelman1980,Wilen1988,Dykman1993b,Kristensen1996,Konstantinov2013,Chepelianskii2015,Rees2016} and references therein. However, the only effect of their interaction on the spectral lines studied so far is a density-dependent small line shift \cite{Lambert1980,Konstantinov2009}.

In this paper we show that, by applying  a magnetic field along the helium surface, one can use the spectroscopy of the electron system to study quantum many-electron dynamics in the liquid and solid states. Importantly, in the cases where this dynamics has been already understood, the system can serve as a quantum simulator of color center spectroscopy, with the unique opportunity of controlling the strength of the coupling of the electron transition and many-body excitations in the system. The importance of such simulations follows from the broad applications of color centers, including the color centers in diamond such as NV centers, cf. \cite{Awschalom2018,Barry2020,Bhaskar2020}.

The change of the interband absorption spectrum by an in-plane magnetic field   has been studied for degenerate quasi-two-dimensional electron systems in semiconductors, see \cite{Ando1982} and references therein.   The results were interpreted in the mean-field approximation. The field-induced high-temperature spectral broadening  was also reported for electrons on helium \cite{Zipfel1976,Zipfel1976a}. Here we show that, for electrons on helium, in a qualitative distinction from the mean-field theory, the effect of the in-plane field is determined by the interplay of the strong correlations and fluctuations in the quantum electron system.

\begin{figure}[h!]
\centering
\includegraphics[width=5.0cm]{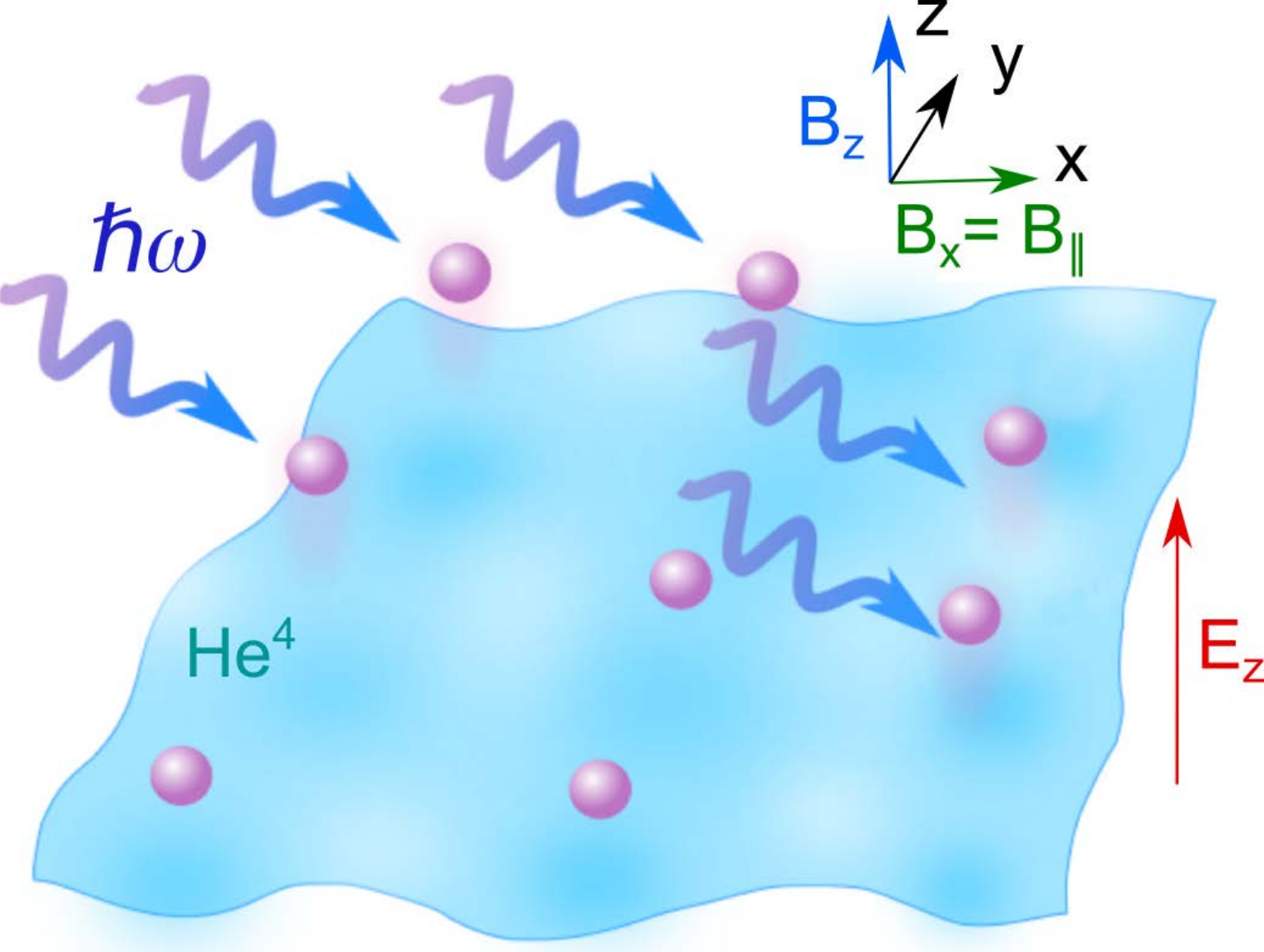} \quad \includegraphics[width=2.6cm]{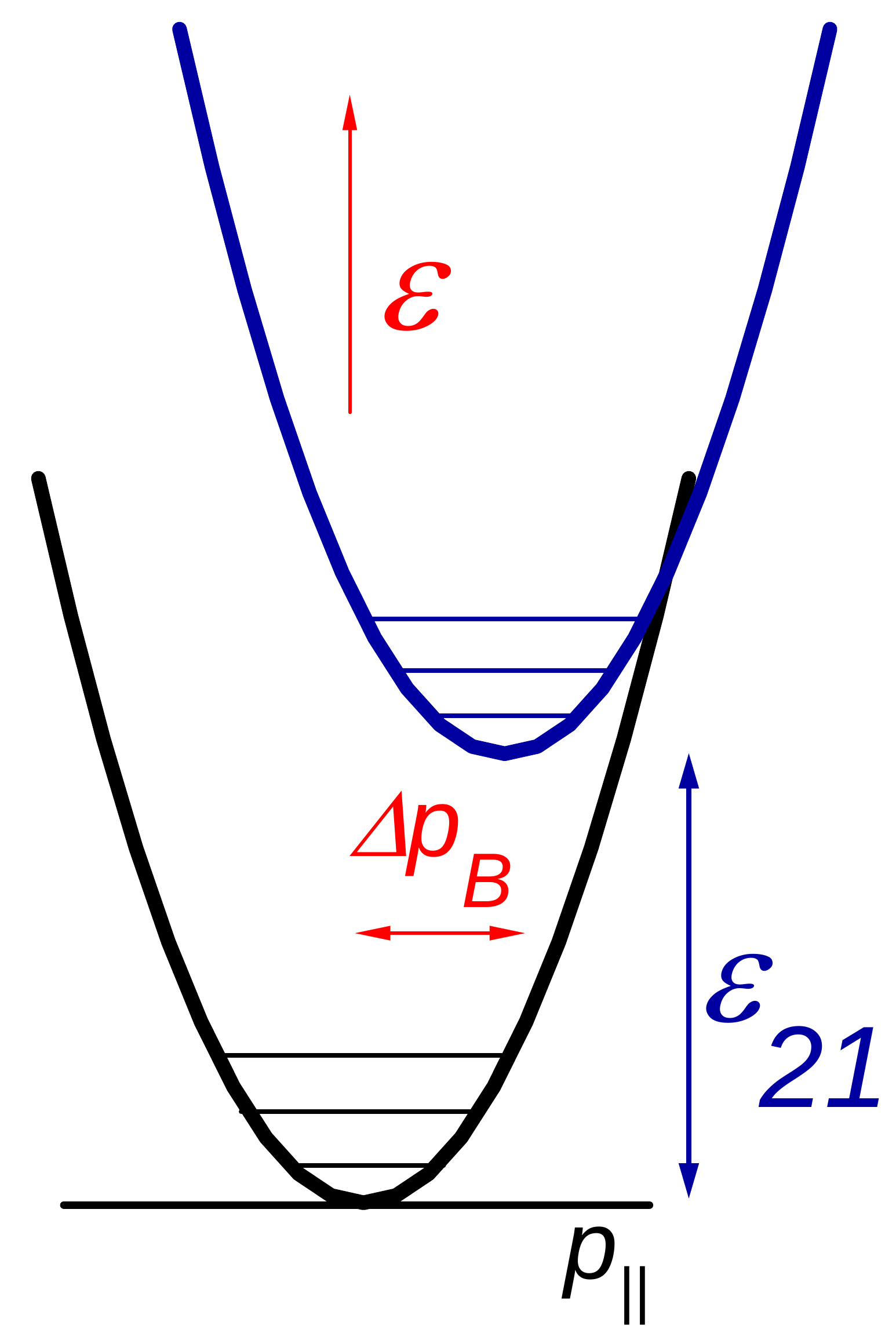} \quad 
\caption{Left: The many-electron system on helium in a magnetic field with components parallel ($\Bp\equiv B_x$) and perpendicular ($\Bperp\equiv B_z$) to the helium surface. Right: The energy spectrum of an electron in the two lowest bands of motion normal to the surface. The energy difference between the bands $\ep_{21}$ is the distance between the levels of the quantized motion along the $z$-axis. The kinetic energy of the electron motion along the surface is quadratic in the in-plane momentum $\pb_\parallel$ for $\Bperp=0$; the field $\Bperp$ transforms the spectrum into discrete Landau levels.  The field $\Bp$  shifts the bands of the in-plane motion by $\Delta p_B$, see Eq.~(\ref{eq:momentum_shift}).}
\label{fig:B_geometry} 
\end{figure} 

The effect of the parallel magnetic field on the electron spectrum  and the similarity with the physics of color centers can be understood from Fig.~\ref{fig:B_geometry}. We choose the $z$ axis as the direction of quantized motion normal to the surface. In different quantized states the electron is at a different average distance from the surface. If a magnetic field $\Bp$ is applied parallel to the surface, an inter-state transition leads to the electron shift transverse to $\Bp$. Therefore the electron in-plane momentum is changed by the Lorentz force in the $\hat{\bf z}\times {\bf B}_\parallel$ direction. For the transition $\Ket{1}\to \Ket{2}$ from the ground to the first excited state the change $\Delta p_B$ is 
\begin{align}
\label{eq:momentum_shift}
&\Delta p_B = m\omp \Delta_z, \qquad \Delta_z = 
\bar{z}_{22} - \bar{z}_{11}, \nonumber\\
&\omp = e\Bp/mc, \qquad \bar{z}_{\mu\mu} =\Bra{\mu}z\Ket{\mu} \quad(\mu=1,2).
\end{align}
Thus the  minima of the energy bands $\ep_1(\pb)$ and $\ep_2(\pb)$ of the in-plane motion ($\pb$ is the in-plane momentum) are shifted with respect to each other. We assume $m\omp^2\Delta_z^2 \ll \ep_{21}\equiv \min[\ep_2(\pb)-\ep_1(\pb)]$.

The right panel of Fig.~\ref{fig:B_geometry} has the familiar form of the sketch of the energy of a point defect  coupled to a vibrational mode in a crystal \cite{Stoneham2001}. In the case of a defect,  the horizontal axis is the coordinate of the vibrational mode, and the parabolas show the potential energy of the mode in the two electron states with  the energy difference $\ep_{21}$. The zero-phonon spectral line corresponds to a transition at frequency $\ep_{21}/\hbar$ between the minima of the parabolas. The vertical transition from the minimum of the lower parabola (the Franck-Condon transition) occurs at a higher energy. Usually the electron is coupled to many modes (phonons), which significantly complicates the analysis, as has been known since the work of Pekar \cite{Pekar1950} and Huang and Rhys \cite{Huang1950}.   

In distinction from  a defect, the parabalae in Fig.~\ref{fig:B_geometry} show the electron energy as a function of the in-plane momentum. In a strongly correlated electron system the momentum can be transferred to other electrons. Such recoil reminds the recoil from the absorption of a gamma-quantum by an impurity in a crystal, which underlies the M\"ossbauer effect. Therefore,  by analogy with the M\"ossbauer effect and the spectra of color centers, the absorption spectrum of electrons on helium should strongly depend on the in-plane electron dynamics. 

%The effect of the in-plane magnetic field  on the interband absorption spectrum has been studied for degenerate quasi-two-dimensional electron systems in semiconductors, see \cite{Ando1982} and references therein.   The results were interpreted in the mean-field approximation. The high-temperature spectral broadening in a field $\Bp\leq 0.2$~T was also reported for electrons on helium \cite{Zipfel1976,Zipfel1976a}. Here we show that, for electrons on helium, in a qualitative distinction from the mean-field theory, the effect of the in-plane field is determined by the interplay of the strong correlations and fluctuations in the nondegenerate quantum electron system. 

The Hamiltonian of the electron system is a sum of the terms $H_\parallel, H_\perp$, and $H_i$ that describe, respectively, the in-plane motion for $\Bp\equiv B_x=0$, the quantized vertical motion in the image charge potential \cite{Andrei1997,Monarkha2004}, and the coupling of these two motions:
\begin{align}
\label{eq:Hamiltonian}
&H=  H_\parallel+ H_\perp + H_i, \quad 
H_\parallel=\sum_n\frac{ {\bm \pi}_n^2}{2m} +\frac{1}{2}\sumprime{n,m}\frac{e^2}{|\rb_n-\rb_m|},
\nonumber\\
&H_\perp =\sum_n \left[\frac{p_{nz}^2}{2m} +U(z_n)\right], \quad
H_i=\sum_n \omp \pi_{ny}(z_n- \bar{z}_{11}).
\end{align}
Here $n$ enumerates electrons, $\rb_n \equiv (x_n,y_n)$ and ${\bm \pi}_n = -i\hbar \n_n+(e/c){\bf A}_\perp(\rb_n)$ are  the in-plane electron coordinate and kinematic momentum [${\bf A}_\perp (\rb)$ is the vector-potential of the field $\Bperp$], whereas $U(z)$ is the confining potential, which includes the term $m\omp^2(z-\bar{z}_{11})^2/2$ due to the field $\Bp$. This term leads to a comparatively small change of the distance $\ep_{21}$ between the energy levels  of the out-of-plane motion. We have omitted  the out-of-plane component of the electron-electron interaction. It makes $\ep_{21}$ weakly density-dependent \cite{Lambert1981,Konstantinov2009}. 
The leading-order part of $H_i$ is diagonal with respect to the states $\Ket{\mu}_n$  of the out-of-plane motion, $
H_i = \omp\Delta_z\sum_n \pi_{ny}  \ket{2}\!_{n\;n}\!\!\bra{2}$, see Supplemental Material (SM) \footnote{The Supplemental Material provides the details of the calculation and the relation to the SI units}.

Absorption of microwaves polarized in the $z$-direction is determined by the real part of the conductivity $\sigma_{zz}(\omega)$. For a nondegenerate electron system it is given by the  sum of the contributions from individual electrons. We write it as the conductivity of an $n_{}$th electron multiplied by the in-plane electron density $n_s$,
\begin{align}
\label{eq:conductivity_general}  
&\mathrm{Re}\,\sigma_{zz}(\omega) =C_\sigma\mathrm{Re}\int_0^\infty dt e^{i\omega t}
\langle z_{n_{}}(t)z_{n_{}}(0)\rangle.
\end{align}
Here $C_\sigma=e^2n_s\omega/\hbar\approx e^2n_s\ep_{21}/\hbar^2$ in the considered range of resonant absorption; $z_n$ is counted off from $\bar z_{11}$.

It is convenient to calculate the time correlator of $z_n$ in the standard interaction representation using the operator $U (t)= \exp[-i(H_\parallel + H_\perp )t/\hbar]$. This gives 
\begin{align}
\label{eq:Q_general}
&\langle z_{n_{}}(t)z_{n_{}}(0)\rangle = \bigl\vert \Bra 1 z \Ket 2\bigr\vert^2 \exp(-i\ep_{21} t/\hbar)Q(t), \nonumber\\
&Q_{}(t)=  \left\langle T_\tau \exp\left[-i(\omp \Delta_z/\hbar)\int_0^t d\tau \pi_{ny}(\tau)\right]\right\rangle.
\end{align}
The notation $\langle\cdot\rangle$ indicates thermal averaging over the in-plane many-electron states in the ground state of the out-of-plane motion; $T_\tau$ is the time-ordering operator.

Function $Q_{}(t)$ determines the shape of the absorption peak near the inter-subband frequency $\ep_{21}/\hbar$. It can be found in the explicit form if the electrons form a Wigner crystal (SM). In this case $\pi_{ny}$ is expressed in terms of the phonon variables of the crystal, making the formulation essentially identical to that in the problem of electron-phonon transitions in color centers.

 In the regime of our experiment the electron system is a strongly correlated liquid in a strong transverse magnetic field. It has magnetoplasmon modes with frequencies $\geq\omega_c$ and low-energy excitations with the bandwidth $\sim \hbar\omega_p^2/\omega_c$, where
\begin{align}
\label{eq:frequencies}
\omega_c \equiv e\Bperp/mc \gg \omega_p^2/\omega_c, \quad \omega_p =(2\pi e^2 n_s^{3/2}/ m)^{1/2}.
\end{align}
For $\hbar\omega_c\gtrsim k_BT$, the in-plane electron motion in the range (\ref{eq:frequencies}) is a superposition of fast quantized cyclotron motion and a slow semiclassical drift of the guiding centers of the cyclotron orbits in the fluctuational electric field caused by the electron density fluctuations \cite{Dykman1979a}. The kinematic momentum in the interaction representation is (see SM)
\begin{align}
\label{eq:kinematic}
\pi_{ny}(t)  \approx [\tilde{\pi}_{ny}(t)e^{i\omega_c t}+\mathrm{H.c.}] + (e/\omega_c)E_{nx}(t).
\end{align}
Here, $\mathbf{E}_n=-e\sum^{\prime}_m(\rb_n-\rb_m)/|\rb_n-\rb_m|^3$ is the fluctuational field on the $n$th electron. It is semiclassical for $k_BT\gg \hbar\omega_p^2/\omega_c$ and $n_s \ell^2\ll 1$ [$\ell=(\hbar/m\omega_c)^{1/2}$]. Both $\mathbf{E}_n(t)$ and ${\tilde{\bm \pi}}_n(t)$ vary on the time scale $\omega_c/\omega_p^2\gg 1/\omega_c$.

Where the cyclotron frequency $\omega_c$ largely exceeds the width of the spectral peak near $\ep_{21}/\hbar$, this peak can be described by averaging $Q_{}(t)$ over time $\sim \omega_c^{-1}$. From Eqs.~(\ref{eq:Q_general}) - (\ref{eq:kinematic}) the resulting function $\bar Q_{}(t)$ has the form
\begin{align}
\label{eq:central_peak}
&\bar Q_{}(t) = e^{i\delta_\parallel t}\exp[-(\gamma^2/2)w(t)],\nonumber\\
&\delta_\parallel =m\omp^2\Delta_z^2/2\hbar, \quad \gamma^2 = \delta_\parallel \omega_p^2k_BT/2\pi\hbar\omega_c^2,\nonumber\\
&w(t) = 
(n_s^{3/2}k_BT)^{-1}\iint_0^t dt_1\,dt_2\, \langle \mathbf{E}_{n}(t_1)\mathbf{E}_{n}(t_2)\rangle.
\end{align}
We assumed Gaussian distribution of the fluctuational field $\mathbf{E}_n$, see SM. In a broad parameter range relevant for the experiments on electrons on helium $\langle \mathbf{E}_n^2\rangle \approx F(\Gamma) n_s^{3/2}k_BT$, where $F(\Gamma) \simeq 9$.

If the coupling to the in-plane fluctuations is strong,  $\gamma\gg \omega_p^2/\omega_c$, from Eq.~(\ref{eq:central_peak})  the main part of the absorption spectrum (\ref{eq:conductivity_general}) is a Gaussian peak,  reminiscent of the spectrum of color center. The typical width of the peak in the frequency units is $\gamma F(\Gamma)^{1/2}$. 

The absorption spectrum also has an analog of the zero-phonon line. It is described by the long-time behavior of $w(t)$ and dominates the spectrum for weak coupling (small $\Bp$). It follows from the analysis that in the electron liquid the line is Lorentzian with a half-width determined by the self-diffusion and equal to $m\delta_\parallel D/2$, where $D$ is the self-diffusion coefficient, see SM. One may expect to switch from a Lorentzian to a Gaussian spectrum by increasing the field $\Bp$.

\begin{figure}[!htb]
\centering
\includegraphics[width=0.9\columnwidth]{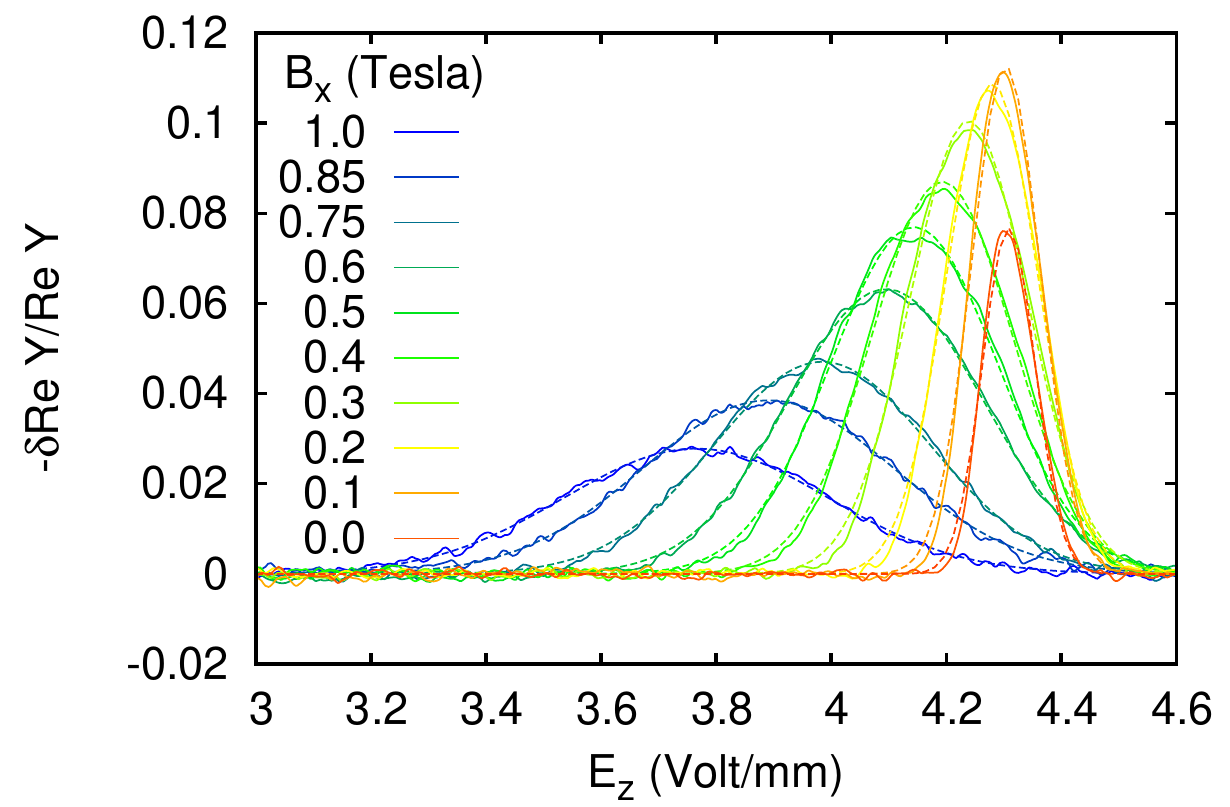}
\caption{The spectra of the relative microwave-induced change of the low-frequency admittance $Y$ for different $\Bp\equiv B_x$. The data refers to the microwave frequency $f = 150$~GHz, $T = 0.2$~K, $\Bperp\equiv B_z = 0.5$~T, and $n_s = 21.5 \times 10^6\;{\rm cm^{-2}}$. The AC bias is 30mV. The dashed lines show Gaussian fit to the data with the variance $\delta E_z^2$ given by Eq.~(\ref{eq:Gaussian_width}). 
}
\label{fig:spectra} 
\end{figure}

In the experiment, the absorption spectrum is measured by varying the  electric field $E_z$ applied perpendicular to the helium surface, using that the level spacing $\ep_{21}$ linearly depends on $E_z$ within the linewidth. In the units of $E_z$, the typical width of the Gaussian peak is 
\begin{align}
  \delta E_z &=  \frac{\Bp}{\Bperp \sqrt{2}} \left[ k_B T n_s^{3/2} F(\Gamma) \right]^{1/2}.
\label{eq:Gaussian_width}
\end{align}
All parameters in Eq.~(\ref{eq:Gaussian_width}) can be controlled in the experiment. This enables testing the theoretical prediction with high accuracy. 

We measured the change of the low-frequency helium cell admittance $Y$ due to absorption of microwave radiation. Such photo-assisted transport spectroscopy provides a sensitive way to measuring resonant microwave absorption \cite{Konstantinov2009}. The method has been used to study the rich out-of-equilibrium physics and a variety of nontrivial nonlinear effects  associated with moderately strong resonant microwave excitation of the electron system \cite{Konstantinov2009,Konstantinov2013,Chepelianskii2015,Yunusova2019}. Here we focus on the linear response. The microwave power was attenuated down to $\mu{\rm W}$ levels. Except for the low power, the experiment was done in the same way as in \cite{Yunusova2019} where the focus was on the single-electron physics.

The spectra of the resonant $\Ket{1}\to \Ket{2}$ photoexcitation are shown in Fig.~\ref{fig:spectra}. For $\Bp\gtrsim 0.4$~Tesla, where the strong-coupling condition holds, the observed shape of the spectra is very well described by a Gaussian fit (dashed lines) with the variance $\delta E_z$ given by Eq.~(\ref{eq:Gaussian_width}), with no fitting parameters. The overall area of the spectral peaks is determined by the photo-assisted transport response of electrons on helium, which depends on $\Bp$; the discussion of this dependence is beyond the scope of this paper.
\begin{figure}[!htb]
\centering
\includegraphics[width=0.9\columnwidth]{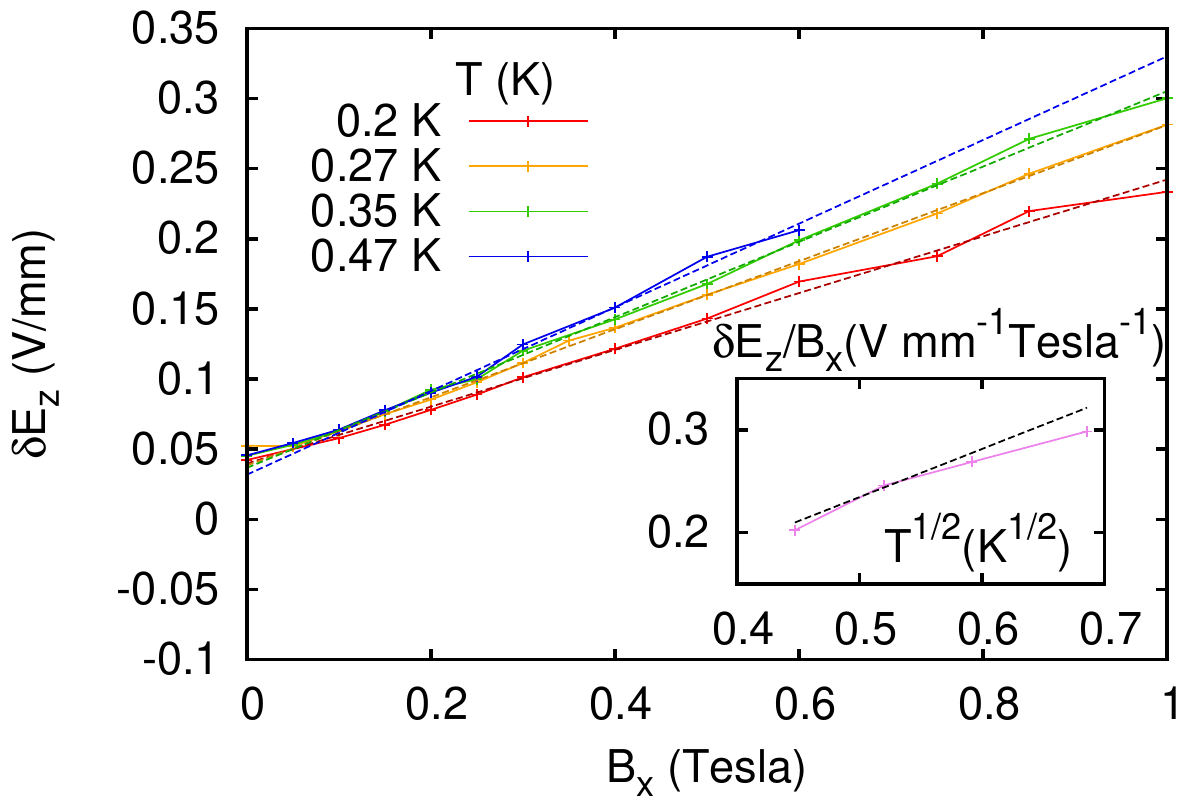}
\caption{The dependence of the typical width of the spectral peaks $\delta E_z$ on $\Bp\equiv B_x$ for different temperatures. The other parameters are the same as in Fig.~\ref{fig:spectra}. The dashed lines are the linear fit. In the strong-coupling range they are described by Eq.~(\ref{eq:Gaussian_width}).  The inset shows the slope of $\delta E_z/\Bp$ as a function of $T^{1/2}$. The dashed black line is given by Eq.~(\ref{eq:Gaussian_width}) with no adjustable parameters. }
\label{fig:different_T}
\end{figure}

In Fig.~\ref{fig:different_T} we show the linewidth $\delta E_z$ as a function of $\Bp$ for several refrigerator temperatures. The observed linear dependence quantitatively agrees with Eq.~(\ref{eq:Gaussian_width}) in the strong-coupling regime, which corresponds to $\Bp T^{1/2} \gtrsim 0.15$~Tesla$\times$K$^{1/2}$, for the used $n_s$ and $\Bperp$. The linewidth at $\Bp = 0$ is attributed to residual inhomogeneous broadening in our system. The linear fits to the data at different temperatures all intersect near $\Bp = 0$, supporting this interpretation. The inset shows the ratio $\delta E_z/\Bp$ as a function of the square root of the temperature. The black line depicts this ratio as given by Eq.~(\ref{eq:Gaussian_width}) with no adjustable parameters [Eq.~(\ref{eq:Gaussian_width}) holds for $T^{1/2}<(\hbar\omega_c/k_B)^{1/2} \approx 0.6$~K$^{1/2}$].

\begin{figure}[!htb]
\centering
\includegraphics[width=1\columnwidth]{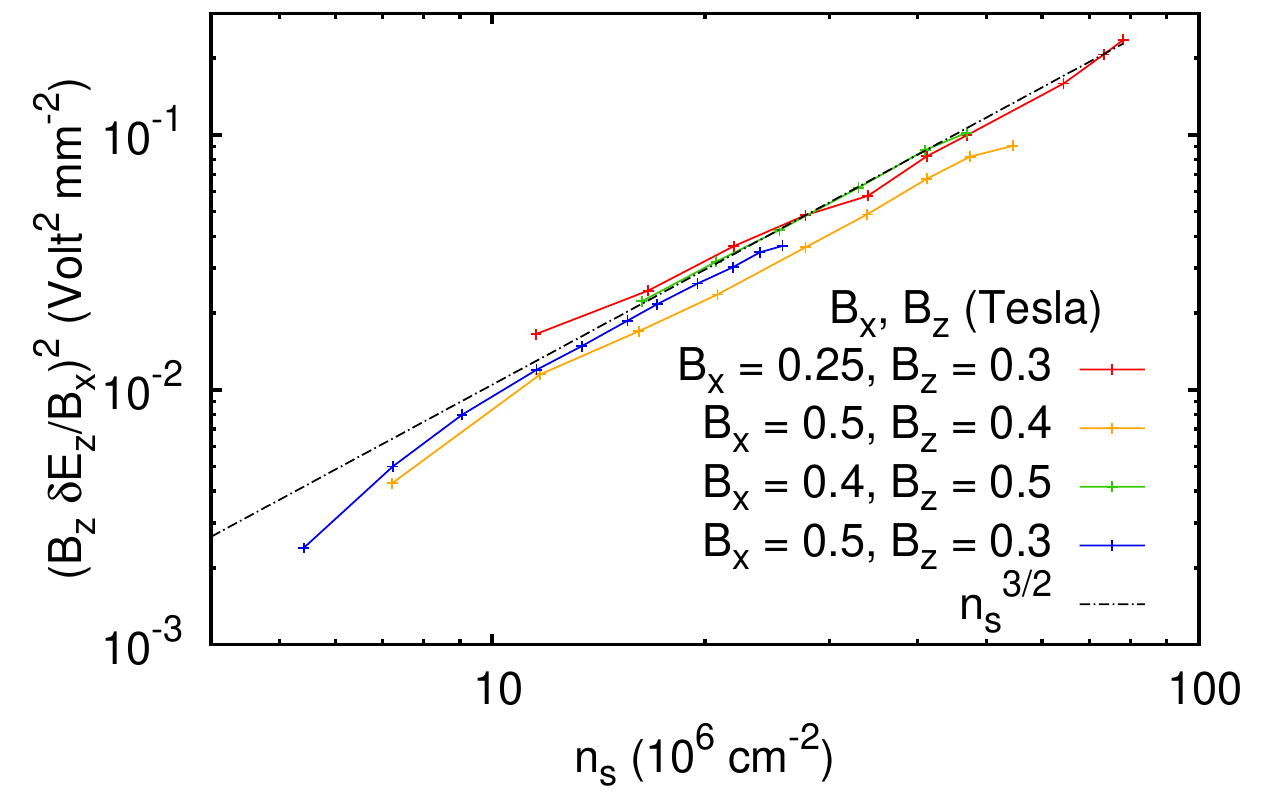}
\caption{The dependence of the squared linewidth scaled by the ratio of the magnetic fields  $\Bp\equiv B_x$ and $\Bperp\equiv B_z$ on the electron density $n_s$. The excitation frequency is $174$~GHz except for the data at $\Bperp\equiv B_z = 0.3$~Tesla, which refers to $f = 144$~GHz. The black line shows the prediction of Eq.~(\ref{eq:Gaussian_width}) for the effective electron temperature $T_e = 0.6$ K.
}
\label{fig:different_density}
\end{figure}

To further check Eq.~(\ref{eq:Gaussian_width}) we investigated the density dependence of the linewidth  for different magnetic fields $\Bp\equiv B_x$ and $\Bperp\equiv B_z$. In order to reduce the averaging time and increase the sensitivity for small $n_s$ we used a stronger microwave power, in the $100\;\mu{\rm W}$ range. This resulted in an additional spectral broadening,  which we attributed to an effective electron temperature $T_e = 0.6\;{\rm K}$  (the refrigerator temperature was $0.3\;{\rm K}$). With this assumption the data are in full agreement with  Eq.~(\ref{eq:Gaussian_width}). As shown in Fig.~\ref{fig:different_density}, $(\delta E_z)^2\propto n_s^{3/2}$.  By rescaling the linewidth, we see that the results for different  $\Bp$ and $\Bperp$ collapse onto the same curve.

Our results demonstrate that the spectra of resonant interband absorption are in full qualitative and quantitative agreement  with the many-electron theory, with no adjustable parameters. The experimental data present the first direct measurement of the fluctuational electric field that an electron is experiencing in a nondegenerate  electron liquid. The explicit theoretical results refer to the range where the in-plane motion is quantized by the magnetic field. This is a nontrivial regime where the quantization helps revealing the many-electron physics and describing it in terms of the dynamics of the guiding centers of the cyclotron orbits. This dynamics is quasiclassical in a sufficiently strong magnetic field, which was used to obtain the explicit expressions for the parameters of the spectrum. 

The results show that, on the one hand, by applying an in-plane magnetic field one can directly study intimate features of the quantum many-electron dynamics of an electron liquid and a Wigner crystal, including self-diffusion in the electron liquid. The regimes other than the one explored here can be also investigated, particularly for the Wigner crystal where the closed-form expression for the spectrum is obtained. On the other hand, the results provide a means for testing the theory of the spectra of color centers in the setting where the effective coupling strength can be varied in situ by varying the in-plane magnetic field. 

\acknowledgments
ADC acknowledges support from ANR JCJC SPINEX. DK is supported by the internal grant from Okinawa Institute of Science and Technology (OIST) Graduate University. MID was supported in part by the Grant DE-SC0020136 funded by the US Department of Energy, Office of Science.

%%%%%%%%%%%%%%%%%%%%%%%%%%%%%%%
%%%%%%%%%%%%%%%%%%%%%%%%%%%%%%

\clearpage
\begin{widetext}

\begin{center}
{\Large Supplemental Material for the paper\\
{\bf Many-electron system on helium and the color center spectroscopy}\\
%\hfill\\
by A. Chepelianskii, D. Konstantinov, and M. I. Dykman}
\end{center}

\end{widetext}

\setcounter{equation}{0}

\section{Position of the spectral line}
\label{sec:energy_shift}

The position of the spectral line depends on the magnetic field $\Bp$. To the leading order, in the single-electron approximation the $\Bp$-dependent term in the confining potential $U(z_n)$ leads to the shift $\ep_{21}\to \ep_{21} + (m\omp^2/2)[\Bra{2}z^2\Ket{2} - \bra{1}z^2\ket{1}-(\bar z_{22})^2 + (\bar z_{11})^2]$, where we have taken into account that  $U(z_n)$ in the main text contains the term $m\omp^2(z-\bar z_{11})^2/2$. We recall for completeness that 
\[\omp =e\Bp/mc, \quad \bar z_{\mu\mu} = \Bra{\mu}z\Ket{\mu} .\]
%
%Interestingly, the term $\Delta_z^2$ in the above expression for the shift cancels the term $\delta_\parallel$ in Eq.~(\ref{eq:central_peak}) of the main text. So, 
On the other hand, for an electron liquid in the semiclassical strong-coupling limit $\gamma\gg \omega_p^2/\omega_c$,  taking into account Eq.~(\ref{eq:central_peak}) of the main text we find that the dependence of the resonant transition frequency  $\omega_\mathrm{res}$ on the field $\Bp$  has the form
\begin{align}
\label{eq:resonance_position}
\omega_\mathrm{res} \approx \hbar^{-1}\ep_{21} +(m\omp^2/2)( \Bra{2}z^2\Ket{2} - \bra{1}z^2\ket{1}-\Delta_z^2).
\end{align}
where $\Delta_z = \bar z_{22} - \bar z_{11}$ \footnote{We are using CGS units. To switch to the SI units, one should set in the expressions we give  $c=1$, replace $e^2\to e^2/4\pi\epsilon_0$ and, in the main text, in Eq.~(\ref{eq:central_peak}) for $w(t)$ and Eq.~(\ref{eq:Gaussian_width}), replace $n_s^{3/2} \to n_s^{3/2}/4\pi\epsilon_0$.}
.

In the analysis we have disregarded the part of $H_i$ that is non-diagonal with respect to the states of out-of-plane motion $\ket{\mu}_n$. This part can be treated by the perturbation theory for the considered low temperatures, $k_BT\ll \ep_{21}$. To the leading order, it leads to a shift of the Landau levels of the in-plane motion. In the interesting case where $\ep_{21}\gg \hbar\omega_c$ this shift has an opposite sign for the levels in the first and the second states of out-of-plane motion. For the $k$th Landau level ($k=0,1,\ldots$) in the out-of-plane state $\ket{1}$ it is $\approx - (\hbar \omega_c/2\ep_{21})m \omp^2 (2k+1) |\bra{1} z\ket{2}|^2$.  The analysis can be extended to the case where the ratio $\hbar\omega_c/\ep_{21}$ is not small.  For completeness we mention that the shift of  $\ep_{21}$ due to the out-of-plane component of the electron-electron interaction is $\sim m\omega_p^2\Delta_z^2$; the full expression is given in \cite{Konstantinov2009}.

\section{Slow variables in the electron liquid in a strong magnetic field}  
\label{sec:liquid}

As shown in the main text, resonant absorption near the frequency $\ep_{21}/\hbar $ is determined by  the time correlation function $Q(t)$ given by Eq.~(\ref{eq:Q_general}) of the main text. The calculation of $Q(t)$ is done differently for an electron liquid and a Wigner solid, with the calculation for a Wigner solid being simpler, see Sec.~\ref{sec:Wigner}. In the both cases the result depends on the interrelation between the frequency $\omp$, the frequencies that determine the electron dynamics 
\begin{align}
\label{eq:SM_frequencies}
\omega_p=(2\pi e^2n_s^{3/2}/m)^{1/2},\qquad \omega_c = e\Bperp/mc,
\end{align}
and the temperature. Here $\omega_p$ is the characteristic short-wavelength plasma frequency of the 2D electron system in the absence of a magnetic field; it is obtained from the standard expression for the long-wavelength plasma frequency as a function of the in-plane wave number (cf.~\cite{Andrei1997}) by setting this wave number equal to $n_s^{1/2}$ . This is the analog of the Debye frequency of the Wigner crystal in the absence of a magnetic field. 

In the opposite limit of a strong magnetic field $\Bperp$, where the cyclotron frequency $\omega_c\gg \omega_p$, the frequency $\omega_p^2/\omega_c$ is the reciprocal time scale for the motion of the guiding centers, see Sec.~\ref{subsec:guiding_centers}; this is also the limiting frequency of the lower branch of phonons in the Wigner crystal in a strong field. We will concentrate on the case that was studied in the experiment in most detail,
\begin{align}
\label{eq:semiclassical}
\omega_c \gtrsim k_BT/\hbar \gg \omega_p^2/\omega_c.
\end{align}

\subsection{Dynamics of the guiding center}
\label{subsec:guiding_centers}
 
In a strong transverse magnetic field, $\omega_c\gg \omega_p$, it is convenient to change to the fast oscillating momentum components of individual electrons $\pi_\pm$ and the  slowly varying guiding center coordinates $R_\pm$, which commute with $\pi_\pm$, 
\begin{align}
\label{eq:pi_pm}
\pi_\pm = (\ell/\hbar\sqrt{2})(\pi_x \mp i\pi_y), \qquad [\pi_-, \pi_+]=1.
\end{align}
Here $\ell=(\hbar c/e\Bperp)^{1/2}$ is the quantum magnetic length (we chose the sign of the magnetic field $\Bperp\equiv B_z$ so that $e\Bperp  > 0$). The guiding center operators are
\begin{align}
\label{eq:guiding_operators} 
&R_\pm = x\mp iy \pm i\ell\sqrt{2}\,\pi_\pm, \qquad [R_\pm, \pi_\pm]=0,\nonumber\\
&[R_+,R_-]=2\ell^2.
\end{align}

The underlying physical picture is that the guiding centers of the electron cyclotron orbits fluctuate about their quasiequilibrium position in the electron liquid or equilibrium positions in the Wigner crystal. The fluctuations are thermal, with the typical mean square displacement $\sim k_BT/m\omega_p^2$. This estimate is obtained by considering a displacement of an electron from  its (quasi)equilibrium position in the field of other electrons, given that this displacement is small compared to the inter-electron distance $\sim n_s^{-1/2}$. Still the displacement largely exceeds the quantum magnetic length $\ell$ in the parameter  range (\ref{eq:semiclassical}). Therefore the dynamics of the guiding centers in this range is semiclassical \cite{Dykman1979a}.  

On the formal side, the electron kinetic energy in terms of the operators $\pi_{n\pm}$ is $\sum_n({\bm\pi}_n^2/2m) \to \hbar\omega_c\sum_n \pi_{n+} \pi_{n-} +$~const. Operators $\pi_{n\pm}$ play the role of the ladder operators with respect to the Landau energy levels. To the leading order in $\ell^2 n_s$, from Eq.~(\ref{eq:guiding_operators}), the equation of motion for $R_{n\pm}$ in the interaction representation [$\dot R_{n\pm}=-(i/\hbar)[R_{n\pm},H_\parallel]$ with $H_\parallel$ given by Eq.~(\ref{eq:Hamiltonian}) of the main text],  reads
\begin{align}
\label{eq:eom_guiding}
&\dot R_{n\pm}= \mp ic\tilde E_{n\pm}/\Bperp, \quad \tilde E_{n\pm}=  
-e\sumprime{m} R_{nm\pm}/|R_{nm+}|^{3},
\nonumber\\
& R_{nm\pm} = R_{n\pm}-R_{m\pm}, \quad \tilde E_{n\pm} \equiv (\tilde E_{nx} \mp i \tilde E_{ny}).
\end{align}
Here we have used that the magnetic length $\ell$ is small compared to the interelectron distance $|\rb_n - \rb_m| \approx |R_{nm\pm}|$. If we disregard corrections $\sim \ell^2n_s$, the field $\tilde{\mathbf{E}}_n$ coincides with the field $\mathbf{E}_n$ used in the main text. In the approximation $\ell^2n_s\ll 1$ we have also disregarded the non-commutativity of the position operators of the guiding centers $R_{n+}$ and $R_{n-}$. This semiclassical approximation breaks down for $k_BT < \hbar\omega_p^2/\omega_c$. The analysis of the low-temperature case can be done assuming that the electrons form a Wigner crystal, see Sec.~\ref{sec:Wigner}. 

With the account taken of the relation $|R_{nm\pm}|\gtrsim n_s^{-1/2}$, one can see from Eq.~(\ref{eq:eom_guiding}) that the time scale on which the guiding orbit centers $R_{n\pm}$ are changing is given by $\omega_c/\omega_p^2$.  The field $\tilde E_{n\pm}$ varies on the same time scale.

To the first order in $\ell n_s^{1/2}$, the equation of motion for the operators $\pi_{n\pm}$ in the interaction representation is
\begin{align}
\label{eq:eom_pi}
&\dot \pi_{n\pm} \equiv -\frac{i}{\hbar}[\pi_{n\pm},H_0]= \pm i\omega_c \pi_{n\pm} -
\frac{\ell}{\hbar\sqrt{2}}eE_{n\pm},\\
&E_{n\pm} \approx \tilde E_{n\pm} \mp i\frac{e\ell}{\sqrt{2}}
\sumprime{m}\left[\frac{\pi_{nm\pm}}{|R_{nm+}|^3}  - 3 \frac{R_{nm\pm}^2\pi_{nm\mp}}{|R_{nm+}|^5}\right], \nonumber
\end{align}
where $\pi_{nm\pm} = \pi_{n\pm}-\pi_{m\pm}$.

It is seen from Eq.~(\ref{eq:eom_pi}) that the operators $\pi_{n\pm}$ oscillate in time as $\exp(\pm i\omega_c t)$. 
The field $\tilde E_{n\pm}$, on the other hand, varies on a much slower time scale $\sim (\omega_p^2/\omega_c)^{-1}$ . 
For $\omega_c\gg \omega_p^2/\omega_c$ we can write the solution of Eq.~(\ref{eq:eom_pi}) in the adiabatic approximation as
\begin{align}
\label{eq:adiabatic}
\pi_{n\pm}(t) \approx \tilde\pi_{n\pm}(t)e^{\pm i\omega_c t} \mp i\frac{e\ell}{\hbar\omega_c\sqrt{2}}\tilde E_{n\pm}(t)
\end{align} 
This expression was used in Eq.~(\ref{eq:kinematic}) of the main text with $
\tilde{\mathbf{E}}_n$ replaced by $\mathbf{E}_n$.

The time dependence of the operators $\tilde\pi_{n\pm}$ is determined, to the lowest order, by the first term in the square bracket in the expression (\ref{eq:eom_pi}) for $E_{n\pm}$. To the zeroth order in the electron-electron interaction, this term is oscillating as $\exp(\pm i\omega_ct)$, i.e., in the same way as $\pi_{n\pm}$, whereas the last term in the expression for $E_{n\pm}$ is counter-rotating, it oscillates as $\exp(\mp i\omega_c t)$. As seen from Eq.~(\ref{eq:eom_pi}), $\tilde\pi_{n\pm}$ varies on the time scale $(\omega_p^2/\omega_c)^{-1}$ \cite{Dykman1979a}.

\subsection{Averaging the expression for the time correlation function}
\label{subsec:averaging}

In the considered parameter range (\ref{eq:semiclassical}), where the dynamics of the guiding centers is semiclassical, in the expression for the correlation function, Eq.~(\ref{eq:Q_general}) of the main text, one can do the averaging over the quantum cyclotron motion and over the positions of the guiding  centers separately. This leads to the expression
\begin{align}
\label{eq:averaging}
&Q_{}(t)\approx I^{(1)}(t)I^{(2)}(t); \qquad I^{(1)}(t)  = \left\langle T_\tau\exp\left[\frac{\omp\Delta_z}{\ell\sqrt{2}} \right.\right.\nonumber\\
& \left.\left.\times\int_0^t d\tau \left(\tilde\pi_{n_{}+}(\tau)e^{i\omega_c \tau}-\tilde\pi_{n_{}-}(\tau)e^{-i\omega_c\tau}\right)\right] 
\right\rangle,\nonumber\\
&I^{(2)}(t)=
 \left\langle T_\tau \exp\left[i\frac{e\omp \Delta_z}{\hbar\omega_c}\int_0^t d\tau \tilde E_{n_{} x}(\tau)\right]\right\rangle.
\end{align}

Function $Q_{}(t)$ should be found for $t$ on the order of the reciprocal width of the absorption spectrum. We will assume that this width is much smaller than $\omega_c$. Respectively, we are interested in the values of $I^{(1,2)}$ for $\omega_c t\gg 1$. To find $I^{(1)}$ in this range one can use the approximation of non-intersecting diagrams. 
Odd-order terms in the series expansion of $I^{(1)}(t)$ in $\tilde\pi_{n\pm}$ vanish: they contain an odd number of the operators $\tilde\pi_{n_{}\pm}$, and their diagonal matrix element on the wave functions of the Landau levels is zero. We can write the $2k$th term in the expansion of $I^{(1)}(t)$ as

\begin{align}
\label{eq:2n_term}
I^{(1)}_k(t)=&\alpha^{2k}\int_{t_2}^t dt_1 \hat f(t_1)\int_{t_3}^t dt_2 \hat f(t_2) \int_0^t dt_3 \hat f(t_3)...\nonumber\\
&\times \int_0^{t_{2k-1}}dt_{2k}\hat f(t_{2k}), \quad \alpha=\omp \Delta_z/\ell\sqrt{2},\nonumber\\
&\hat f(t) = \tilde\pi_{n+}e^{i\omega_c t}-\tilde\pi_{n-}e^{-i\omega_ct}
\end{align}
Integration over $dt_1$ gives terms that contain fast-oscillating factors $\exp(\pm i\omega_ct)$ and $ \exp(\pm i\omega_c t_2)$. When multiplied by $\hat f(t_2)$, the first group of terms will be fast oscillating as $\exp[\pm i\omega_c(t\pm t_2)]$. Their integral over $t_2$ will be $\propto \omega_c^{-1}$. In contrast, the second group of terms, when multiplied by $\hat f(t_2)$, will lead to the onset of smooth terms. Integrating them over $t_2$ will give a factor $\propto t-t_3\gg \omega_c^{-1}$,  
\begin{align*}
&\int_{t_2}^t dt_1 \hat f(t_1)\int_{t_3}^t dt_2 \hat f(t_2)\\
&\approx \frac{i}{\omega_c}\int_{t_3}^t dt_2 [\tilde\pi_{n-}(t_2),\tilde\pi_{n+}(t_2)]=\frac{i}{\omega_c}(t-t_3).
\end{align*}

Substituting this expression into Eq.~(\ref{eq:2n_term}), differentiating over $t$, and integrating the resulting chain of equations for $I^{(1)}_k(t)$, we obtain:
\begin{align}
\label{eq:2n_term_derivative}
\frac{dI^{(1)}_k}{dt} = (i\alpha^2/\omega_c) I^{(1)}_{k-1}, \qquad I^{(1)}_k =(i\alpha^2t/\omega_c)^k/k!
\end{align}
This gives, to the leading order in $\alpha^2$,
\begin{align}
\label{eq:I1}
I^{(1)}(t) = \exp(i\delta_\parallel t), \quad \delta_\parallel = \omp^2\Delta_z^2/2\ell^2\omega_c,
\end{align}
which is the expression used in Eq.~(\ref{eq:central_peak}) of the main text to describe the contribution of the oscillating terms in $\pi_{ny}$.

The term $I^{(2)}(t)$ can be calculated assuming that the distribution of the fluctuational field in the electron liquid is Gaussian. This assumption was used in obtaining Eq.~(\ref{eq:central_peak}) of the main text. Numerical simulations \cite{FangYen1997} have shown that the single-time distribution of the fluctuational field is indeed very close to Gaussian except for very far tails. The Gaussian distribution is to be expected, since the electron dynamics is primarily random harmonic vibrations about the (quasi)equilibrium position in the electron liquid or a Wigner crystal. This argument also suggests that  the probability density functional of the fluctuational field is Gaussian on the time scale $\gtrsim \omega_c/\omega_p^2$.

 We note that, with the account taken of Eq.~(\ref{eq:eom_guiding}), it is convenient to write the integral of the fluctuational field correlator as 
\begin{align}
\label{eq:field_correlator}
&\iint_0^t dt_1 dt_2\langle \Eb_n(t_1)\Eb_n(t_2)\rangle \approx  \iint_0^t dt_1 dt_2 \langle \tilde\Eb_n(t_1)\tilde\Eb_n(t_2)\rangle\nonumber\\
& =(\Bperp/c)^2\iint_0^t dt_1 dt_2\langle \dot \Rb_n(t_1)\dot \Rb_n(t_2)\rangle \nonumber\\
&=(\Bperp/c)^2\langle [\Rb_n(t)-\Rb_n(0)]^2\rangle. 
\end{align}
This expression relates $I^{(2)}(t)$, and thus the integral of the correlation function of the fluctuational field in Eq.~(\ref{eq:central_peak}) of the main text, to the mean square displacement of a guiding center.

On the time scale much longer than $\omega_c/\omega_p^2$, if electrons form a liquid, they are diffusing. Self-diffusion involves correlated many-electron  motion, as seen in the simulations \cite{Moskovtsev2019}. Therefore it is reasonable to assume that the distribution of the diffusion trajectories is Gaussian. We can then again do the averaging over the fluctuational field in $I^{(2)}(t)$  assuming the field distribution to be Gaussian. From Eq.~(\ref{eq:field_correlator}), on  a long time scale, the leading-order term in the integral of the correlation function of the fluctuational field in Eq.~(\ref{eq:central_peak}) of the main text  becomes %with the account taken of Eq.~(\ref{eq:eom_guiding}) can be written as 
\begin{align}
\label{eq:diffusion}
&\iint_0^t dt_1 dt_2\langle \Eb_n(t_1)\Eb_n(t_2)\rangle \approx  2(\Bperp/c)^2Dt
\end{align} 
where $D$ is the self-diffusion coefficient. Equation(\ref{eq:diffusion}) was used in the main text to describe the long-time limit of the correlator $Q(t)$ and thus the many-electron analog of the zero-phonon line in the electron liquid.

\section{Wigner crystal}
\label{sec:Wigner}

In the case of a Wigner crystal, we write the Hamiltonian $H_\parallel$ as
\begin{align}
\label{eq:WC_Hamiltonian}
H_\parallel=\hbar \sum_{\kb \nu}\omega_{\kb\nu}a^\dagger_{\kb\nu}a_{\kb\nu}.
\end{align}
Here, $\kb$ is the wave vector of a phonon of the crystal and $\nu=1,2$ is the phonon branch. Operators $a^\dagger_{\kb\nu}$ and $a_{\kb\nu}$ are the phonon creation and annihilation operators and $\omega_{\kb \nu}$ is the phonon frequency. The operator of the kinematic momentum of an $n$th electron is
\begin{align}
\label{eq:p_vs_phonons}
{\bm\pi}_{n_{}}=-im\sum_{\kb\nu}e^{i\kb \Xb_{n_{}}}\omega_{\kb\nu}{\bf A}_{\kb\nu} a_{\kb\nu} +{\rm H.c.}
\end{align}
where $\Xb_n$ is the lattice site position.  The coefficients ${\bf A}_{\kb\nu}\propto (n_sS)^{-1/2}$ ($S$ is the area of the system) give the electron displacement in terms of $a_{\kb\nu}, a^\dagger_{\kb\nu}$. For the Wigner crystal in a strong magnetic field $\Bperp$ they  were obtained in Ref.~\onlinecite{Ulinich1979}. 

Substituting Eq.~(\ref{eq:p_vs_phonons}) into the expression (\ref{eq:Q_general}) of the main text for the correlator $Q_{}(t)$ and calculating the trace over the phonons in a standard way, we obtain
\begin{align}
\label{eq:WC_general}
Q_{}(t) = &\exp\left[-(m\omp^2\Delta_z^2/\hbar\omega_c)g(t)\right],\nonumber\\
g(t)=&\frac{1}{2\ell^2}\sum_{\kb\nu}|{\bf A}_{\kb\nu}|^2 \left[i\left(\sin\omega_{\kb\nu}t - \omega_{\kb\nu} t\right) \right.\nonumber\\
&\left.+ (2\bar n_{\kb\nu}+1)(1-\cos\omega_{\kb\nu}t)\right],
\end{align}
where $\bar n_{\kb\nu} \equiv \bar n(\omega_{\kb\nu})$ is the phonon Planck number, $\bar n(\omega)=[\exp(\hbar\omega/k_BT)-1]^{-1}$. Equation (\ref{eq:WC_general}) is not limited to the case of a strong magnetic field $\Bperp$. It also applies for an arbitrary temperature as long as the electrons form a Wigner crystal. 

Before discussing other limiting cases we show that Eq.~(\ref{eq:WC_general}) coincides with Eq.~(\ref{eq:central_peak}) of the main text  in the strong magnetic field $\Bperp$, when $\omega_c\gg \omega_p$, and when $k_BT\gg \hbar\omega_p^2/\omega_c$ [except that Eq.~(\ref{eq:WC_general}) does not describe electron diffusion in the liquid phase]. We note first that, for a strong field $\Bperp$, the phonon spectrum of the Wigner crystal consists of a high-frequency magnetoplasmon branch $\nu=1$ and a low-frequency branch $\nu=2$. The widths of the both branches are $\sim \omega_p^2/\omega_c$. The branch $\nu=1$ is an analog of the optical phonon branch, $\omega_{\kb\,\nu=1}\to \omega_c$ for $k\to 0$, whereas $\omega_{\kb\,\nu=2}\propto k^{3/2}$ for $k\to 0$. 

The contribution of the branch $\nu=1$ to $g(t)$ consists of fast-oscillating terms $\propto \exp(\pm i\omega_c t)$, which make a small contribution to the smooth part of $Q(t)$, and also of a non-oscillating term  $\propto t$.  Using the results \cite{Ulinich1979}, one can show that $|A_{\kb\,1}|^2 \approx \ell^2/n_sS$, to the leading order in $\omega_p/\omega_c$.  Therefore the term $\propto t$ in $g(t)$ is $\approx -i\omega_c t/2$ and its effect on $Q(t)$ is described by the factor $\exp(i\delta_\parallel t)$ in Eq.~(\ref{eq:central_peak}) of the main text; the correction to $\delta_\parallel$ from the branch $\nu=2$ can be shown to be $\propto \omega_p^2/\omega_c^2\ll 1$.

The branch $\nu=2$ corresponds to vibrations of the electron guiding centers. For $k_BT\gg \hbar\omega_p^2/\omega_c$ these vibrations are semiclassical and are described by Eq.~(\ref{eq:eom_guiding}) linearized in $\Rb_n-\Xb_n$. The mean-square electron displacement is  $\langle(\Rb_n-\Xb_n)^2\rangle \gg \ell^2$, it is determined by the branch $\nu=2$ if we neglect corrections $\sim \ell^2$. Writing  
\[\Rb_n-\Xb_n = \sum_{\kb}e^{i\kb \Xb_{n_{}}}{\bf A}_{\kb\,\nu=2} a_{\kb\,\nu=2} +{\rm H.c.},\]
we obtain from Eq.~(\ref{eq:field_correlator})
\begin{align*}
&\iint^t dt_1 dt_2\langle \Eb_n(t_1)\Eb_n(t_2)\rangle \approx (\Bperp/c)^2\\
&\times\sum_{\kb}|{\bf A}_{\kb\,\nu=2}|^2 (2\bar n_{\kb\,\nu=2}+1)(1-\cos\omega_{\kb\,\nu=2}t)
\end{align*}
(strictly speaking, we should have replaced $2\bar n_{\kb\,\nu=2}+1\to 2k_BT/\hbar\omega_{\kb\,\nu=2}$). This expression has the same form as the corresponding term in Eq.~(\ref{eq:WC_general}). It shows that, indeed, Eq.~(\ref{eq:central_peak}) of the main text coincides with Eq.~(\ref{eq:WC_general}) where electrons form a Wigner crystal.

A somewhat unexpected result follows from Eq.~(\ref{eq:WC_general}) in the classical limit, $k_BT\gg \omega_{\kb\nu}$. It follows from the sum rule $\sum_{\kb \nu}|{\bf A}_{\kb\,\nu}|^2\omega_{\kb\nu} =\hbar/m$ that
\begin{align}
\label{eq:classical}
 g(t) \approx (\omega_c k_BT/2\hbar)t^2, \qquad \omega_{\kb\nu} t\ll 1.
\end{align}
Then from Eq.~(\ref{eq:WC_general}), $Q(t)\approx \exp[-(m\omp^2\Delta_z^2 k_BT t^2/2\hbar^2]$ and the spectrum $\sigma_{zz}(\omega)$ is Gaussian with typical width 
\[\gamma_G^\mathrm{classic} = (m\omp^2\Delta_z^2 k_BT /\hbar^2)^{1/2}.\]
The expansion where we keep only the term $\propto t^2$ in $g(t)$ applies provided $\gamma_G^\mathrm{classic} \gg \max \omega_{\kb\nu}$. 

Interestingly, the same result for the absorption spectrum follows from the general expression (\ref{eq:conductivity_general}) and (\ref{eq:Q_general}) of the main text if one assumes that there is no electron-electron interaction and no transverse magnetic field $\Bperp$, so that $\pi_{ny}$ is independent of time and the momentum distribution is of the Maxwell-Boltzmann form.  This corresponds to the Doppler broadening of the absorption spectrum due to the thermal distribution of the electron momentum. 

Even for a strong in-plane magnetic field $\Bp$ and high temperatures, where Eq.~(\ref{eq:classical}) applies, the many-electron interaction leads to the onset of an analog of the zero-phonon line. For a Wigner crystal this line has zero width. This means that, for low temperatures, the width of this line in the system of electrons on helium is determined by the electron scattering by ripplons and phonons; for $T\gtrsim 0.7$~K it is determined by scattering by the helium vapor atoms \cite{Andrei1997,Monarkha2004}.

\section{Estimation of the factor $F(\Gamma)$ for a Wigner crystal}
\label{sec:FGamma}

As indicated in the main text, the mean square fluctuational field that drives an electron due to the density fluctuations has the form $\langle \mathbf{E}_n^2\rangle = F(\Gamma)n_s^{3/2}k_BT$. The factor $F(\Gamma)$ in this expression can be explicitly calculated if the electrons form a Wigner crystal. It is given by the lattice sum \cite{Dykman1979a,FangYen1997}
\begin{align}
\label{eq:Lattice_sum}
F(\Gamma) = n_s^{-3/2}\sumprime{m}|\mathbf{X}_m - \mathbf{X}_0|^{-3}.
\end{align}
For a hexagonal lattice the lattice sites are $\mathbf{X}_m = (2n_s)^{-1/2} 3^{-1/4}[(2m_1 + m_2)\hat{\bf x} + m_2\sqrt{3}\hat{\bf y}]$ with integer $m_{1,2}$ and with $ \hat{\bf x}$ and $\hat{\bf y}$ being orthogonal unit vectors. Then
\begin{align}
\label{eq:sum_explicit}
F(\Gamma) = 2^{3/2}3^{3/4}\sumprime{{m_1,m_2}}[(2m_1+m_2)^2 + 3m_2^2]^{-3/2},
\end{align}
 where the sum over $m_1,m_2$ runs from $-\infty$ to $\infty$ and the prime indicates that $m_1^2+m_2^2>0$.
 
In Refs.~\cite{Dykman1979a,FangYen1997} the lattice sum (\ref{eq:sum_explicit}) was calculated by summing over a few terms with $(2m_1+m_2)^2 +3 m_2^2\leq K$, integrating over $dm_1dm_2$ for larger $|m_{1,2}|$, and checking that the result weakly depended on $K$. A different way of calculating the sum is to write it as  
\begin{align}
\label{eq:Alexei_sum}
&F(\Gamma)= 2^{1/2}3^{3/4} \sumprime{{m,n}} \frac{1 + (-1)^{m+n}}{( n^2+ 3 m^2)^{3/2} }\nonumber \\
&= 2\frac{12^{3/4}}{\sqrt{\pi}} \int_0^{\infty} \sumprime{m,n} [1 + (-1)^{m+n}] e^{-(n^2 + 3 m^2) x^2} x^2 dx\nonumber\\
&= 2\frac{12^{3/4}}{\sqrt{\pi}} \int_0^{\infty} \left[\theta_3\left(0,e^{-x^2}\right) \theta_3\left(0,e^{-3x^2}\right)\right.\nonumber\\
&\left. + \theta_4\left(0,e^{-x^2}\right)\theta_4\left(0,e^{-3x^2}\right) - 2\right] x^2 dx.
\end{align}
Here we are using the theta functions, 
\begin{align*}
&\theta_3(0,q) = \sum_{n = -\infty}^{\infty} q^{n^2}, \quad
\theta_4(0,q) = \sum_{n = -\infty}^{\infty} (-1)^n q^{n^2}. 
\end{align*}
The advantagious feature of Eq.~(\ref{eq:Alexei_sum}) is that it expresses the lattice sum in terms of the standard functions. It gives $F(\Gamma) \approx 8.893$.

%\bibliography{c:/Users/Mark/Dropbox/Aaa/BibTex/Zotero19}

\begin{thebibliography}{42}%
\makeatletter
\providecommand \@ifxundefined [1]{%
 \@ifx{#1\undefined}
}%
\providecommand \@ifnum [1]{%
 \ifnum #1\expandafter \@firstoftwo
 \else \expandafter \@secondoftwo
 \fi
}%
\providecommand \@ifx [1]{%
 \ifx #1\expandafter \@firstoftwo
 \else \expandafter \@secondoftwo
 \fi
}%
\providecommand \natexlab [1]{#1}%
\providecommand \enquote  [1]{``#1''}%
\providecommand \bibnamefont  [1]{#1}%
\providecommand \bibfnamefont [1]{#1}%
\providecommand \citenamefont [1]{#1}%
\providecommand \href@noop [0]{\@secondoftwo}%
\providecommand \href [0]{\begingroup \@sanitize@url \@href}%
\providecommand \@href[1]{\@@startlink{#1}\@@href}%
\providecommand \@@href[1]{\endgroup#1\@@endlink}%
\providecommand \@sanitize@url [0]{\catcode `\\12\catcode `\$12\catcode
  `\&12\catcode `\#12\catcode `\^12\catcode `\_12\catcode `\%12\relax}%
\providecommand \@@startlink[1]{}%
\providecommand \@@endlink[0]{}%
\providecommand \url  [0]{\begingroup\@sanitize@url \@url }%
\providecommand \@url [1]{\endgroup\@href {#1}{\urlprefix }}%
\providecommand \urlprefix  [0]{URL }%
\providecommand \Eprint [0]{\href }%
\providecommand \doibase [0]{https://doi.org/}%
\providecommand \selectlanguage [0]{\@gobble}%
\providecommand \bibinfo  [0]{\@secondoftwo}%
\providecommand \bibfield  [0]{\@secondoftwo}%
\providecommand \translation [1]{[#1]}%
\providecommand \BibitemOpen [0]{}%
\providecommand \bibitemStop [0]{}%
\providecommand \bibitemNoStop [0]{.\EOS\space}%
\providecommand \EOS [0]{\spacefactor3000\relax}%
\providecommand \BibitemShut  [1]{\csname bibitem#1\endcsname}%
\let\auto@bib@innerbib\@empty
%</preamble>
\bibitem [{\citenamefont {Andrei}(1997)}]{Andrei1997}%
  \BibitemOpen
  \bibinfo {editor} {\bibfnamefont {E.}~\bibnamefont {Andrei}},\ ed.,\
  \href@noop {} {\emph {\bibinfo {title} {Two-{{Dimensional Electron Systems}}
  on {{Helium}} and {{Other Cryogenic Surfaces}}}}}\ (\bibinfo  {publisher}
  {{Kluwer Academic}},\ \bibinfo {year} {Dordrecht, 1997})\BibitemShut
  {NoStop}%
\bibitem [{\citenamefont {{Monarkha, Y.}}\ and\ \citenamefont {{Kono,
  K.}}(2004)}]{Monarkha2004}%
  \BibitemOpen
  \bibfield  {author} {\bibinfo {author} {\bibnamefont {{Monarkha, Y.}}}\ and\
  \bibinfo {author} {\bibnamefont {{Kono, K.}}},\ }\href@noop {} {\emph
  {\bibinfo {title} {Two-{{Dimensional Coulomb Liquids}} and {{Solids}}}}}\
  (\bibinfo  {publisher} {{Springer}},\ \bibinfo {year} {Berlin,
  2004})\BibitemShut {NoStop}%
\bibitem [{\citenamefont {Grimes}\ \emph {et~al.}(1976)\citenamefont {Grimes},
  \citenamefont {Brown}, \citenamefont {Burns},\ and\ \citenamefont
  {Zipfel}}]{Grimes1976a}%
  \BibitemOpen
  \bibfield  {author} {\bibinfo {author} {\bibfnamefont {C.~C.}\ \bibnamefont
  {Grimes}}, \bibinfo {author} {\bibfnamefont {T.~R.}\ \bibnamefont {Brown}},
  \bibinfo {author} {\bibfnamefont {M.~L.}\ \bibnamefont {Burns}},\ and\
  \bibinfo {author} {\bibfnamefont {C.~L.}\ \bibnamefont {Zipfel}},\ }\bibfield
   {title} {\bibinfo {title} {Spectroscopy of {{Electrons}} in
  {{Image}}-{{Potential}}-{{Induced Surface States Outside Liquid Helium}}},\
  }\href@noop {} {\bibfield  {journal} {\bibinfo  {journal} {Phys Rev B}\
  }\textbf {\bibinfo {volume} {13}},\ \bibinfo {pages} {140} (\bibinfo {year}
  {1976})}\BibitemShut {NoStop}%
\bibitem [{\citenamefont {Stern}(1978)}]{Stern1978}%
  \BibitemOpen
  \bibfield  {author} {\bibinfo {author} {\bibfnamefont {F.}~\bibnamefont
  {Stern}},\ }\bibfield  {title} {\bibinfo {title} {Image {{Potential}} near a
  {{Gradual Interface}} between {{Two Dielectrics}}},\ }\href
  {https://doi.org/10.1103/PhysRevB.17.5009} {\bibfield  {journal} {\bibinfo
  {journal} {Phys Rev B}\ }\textbf {\bibinfo {volume} {17}},\ \bibinfo {pages}
  {5009} (\bibinfo {year} {1978})}\BibitemShut {NoStop}%
\bibitem [{\citenamefont {Lambert}\ and\ \citenamefont
  {Richards}(1980)}]{Lambert1980}%
  \BibitemOpen
  \bibfield  {author} {\bibinfo {author} {\bibfnamefont {D.~K.}\ \bibnamefont
  {Lambert}}\ and\ \bibinfo {author} {\bibfnamefont {P.~L.}\ \bibnamefont
  {Richards}},\ }\bibfield  {title} {\bibinfo {title} {Measurement of {{Local
  Disorder}} in a {{Two}}-{{Dimensional Electron Fluid}}},\ }\href@noop {}
  {\bibfield  {journal} {\bibinfo  {journal} {Phys Rev Lett}\ }\textbf
  {\bibinfo {volume} {44}},\ \bibinfo {pages} {1427} (\bibinfo {year}
  {1980})}\BibitemShut {NoStop}%
\bibitem [{\citenamefont {Rama~Krishna}\ and\ \citenamefont
  {Whaley}(1988)}]{RamaKrishna1988}%
  \BibitemOpen
  \bibfield  {author} {\bibinfo {author} {\bibfnamefont {M.~V.}\ \bibnamefont
  {Rama~Krishna}}\ and\ \bibinfo {author} {\bibfnamefont {K.~B.}\ \bibnamefont
  {Whaley}},\ }\bibfield  {title} {\bibinfo {title} {Excess-{{Electron Surface
  States}} of {{Helium Clusters}}},\ }\href
  {https://doi.org/10.1103/PhysRevB.38.11839} {\bibfield  {journal} {\bibinfo
  {journal} {Phys Rev B}\ }\textbf {\bibinfo {volume} {38}},\ \bibinfo {pages}
  {11839} (\bibinfo {year} {1988})}\BibitemShut {NoStop}%
\bibitem [{\citenamefont {Cheng}\ \emph {et~al.}(1994)\citenamefont {Cheng},
  \citenamefont {Cole},\ and\ \citenamefont {Cohen}}]{Cheng1994}%
  \BibitemOpen
  \bibfield  {author} {\bibinfo {author} {\bibfnamefont {E.}~\bibnamefont
  {Cheng}}, \bibinfo {author} {\bibfnamefont {M.~W.}\ \bibnamefont {Cole}},\
  and\ \bibinfo {author} {\bibfnamefont {M.~H.}\ \bibnamefont {Cohen}},\
  }\bibfield  {title} {\bibinfo {title} {Binding of {{Electrons}} to the
  {{Surface}} of {{Liquid Helium}}},\ }\href@noop {} {\bibfield  {journal}
  {\bibinfo  {journal} {Phys Rev B}\ }\textbf {\bibinfo {volume} {50}},\
  \bibinfo {pages} {1136} (\bibinfo {year} {1994})}\BibitemShut {NoStop}%
\bibitem [{\citenamefont {Collin}\ \emph {et~al.}(2002)\citenamefont {Collin},
  \citenamefont {Bailey}, \citenamefont {Fozooni}, \citenamefont {Frayne},
  \citenamefont {Glasson}, \citenamefont {Harrabi}, \citenamefont {Lea},\ and\
  \citenamefont {Papageorgiou}}]{Collin2002}%
  \BibitemOpen
  \bibfield  {author} {\bibinfo {author} {\bibfnamefont {E.}~\bibnamefont
  {Collin}}, \bibinfo {author} {\bibfnamefont {W.}~\bibnamefont {Bailey}},
  \bibinfo {author} {\bibfnamefont {P.}~\bibnamefont {Fozooni}}, \bibinfo
  {author} {\bibfnamefont {P.~G.}\ \bibnamefont {Frayne}}, \bibinfo {author}
  {\bibfnamefont {P.}~\bibnamefont {Glasson}}, \bibinfo {author} {\bibfnamefont
  {K.}~\bibnamefont {Harrabi}}, \bibinfo {author} {\bibfnamefont {M.~J.}\
  \bibnamefont {Lea}},\ and\ \bibinfo {author} {\bibfnamefont {G.}~\bibnamefont
  {Papageorgiou}},\ }\bibfield  {title} {\bibinfo {title} {Microwave
  {{Saturation}} of the {{Rydberg States}} of {{Electrons}} on {{Helium}}},\
  }\href@noop {} {\bibfield  {journal} {\bibinfo  {journal} {Phys Rev Lett}\
  }\textbf {\bibinfo {volume} {89}},\ \bibinfo {pages} {245301} (\bibinfo
  {year} {2002})}\BibitemShut {NoStop}%
\bibitem [{\citenamefont {Degani}\ \emph {et~al.}(2005)\citenamefont {Degani},
  \citenamefont {Farias},\ and\ \citenamefont {Peeters}}]{Degani2005}%
  \BibitemOpen
  \bibfield  {author} {\bibinfo {author} {\bibfnamefont {M.~H.}\ \bibnamefont
  {Degani}}, \bibinfo {author} {\bibfnamefont {G.~A.}\ \bibnamefont {Farias}},\
  and\ \bibinfo {author} {\bibfnamefont {F.~M.}\ \bibnamefont {Peeters}},\
  }\bibfield  {title} {\bibinfo {title} {Bound {{States}} and {{Lifetime}} of
  an {{Electron}} on a {{Bulk Helium Surface}}},\ }\href@noop {} {\bibfield
  {journal} {\bibinfo  {journal} {Phys Rev B}\ }\textbf {\bibinfo {volume}
  {72}},\ \bibinfo {pages} {125408} (\bibinfo {year} {2005})}\BibitemShut
  {NoStop}%
\bibitem [{\citenamefont {Konstantinov}\ \emph {et~al.}(2009)\citenamefont
  {Konstantinov}, \citenamefont {Dykman}, \citenamefont {Lea}, \citenamefont
  {Monarkha},\ and\ \citenamefont {Kono}}]{Konstantinov2009}%
  \BibitemOpen
  \bibfield  {author} {\bibinfo {author} {\bibfnamefont {D.}~\bibnamefont
  {Konstantinov}}, \bibinfo {author} {\bibfnamefont {M.~I.}\ \bibnamefont
  {Dykman}}, \bibinfo {author} {\bibfnamefont {M.~J.}\ \bibnamefont {Lea}},
  \bibinfo {author} {\bibfnamefont {Y.}~\bibnamefont {Monarkha}},\ and\
  \bibinfo {author} {\bibfnamefont {K.}~\bibnamefont {Kono}},\ }\bibfield
  {title} {\bibinfo {title} {Resonant {{Correlation}}-{{Induced Optical
  Bistability}} in an {{Electron System}} on {{Liquid Helium}}},\ }\href
  {https://doi.org/10.1103/PhysRevLett.103.096801} {\bibfield  {journal}
  {\bibinfo  {journal} {Phys Rev Lett}\ }\textbf {\bibinfo {volume} {103}},\
  \bibinfo {pages} {096801} (\bibinfo {year} {2009})}\BibitemShut {NoStop}%
\bibitem [{\citenamefont {Dykman}\ \emph {et~al.}(2017)\citenamefont {Dykman},
  \citenamefont {Kono}, \citenamefont {Konstantinov},\ and\ \citenamefont
  {Lea}}]{Dykman2017}%
  \BibitemOpen
  \bibfield  {author} {\bibinfo {author} {\bibfnamefont {M.~I.}\ \bibnamefont
  {Dykman}}, \bibinfo {author} {\bibfnamefont {K.}~\bibnamefont {Kono}},
  \bibinfo {author} {\bibfnamefont {D.}~\bibnamefont {Konstantinov}},\ and\
  \bibinfo {author} {\bibfnamefont {M.~J.}\ \bibnamefont {Lea}},\ }\bibfield
  {title} {\bibinfo {title} {Ripplonic {{Lamb Shift}} for {{Electrons}} on
  {{Liquid Helium}}},\ }\href {https://doi.org/10.1103/PhysRevLett.119.256802}
  {\bibfield  {journal} {\bibinfo  {journal} {Phys Rev Lett}\ }\textbf
  {\bibinfo {volume} {119}},\ \bibinfo {pages} {256802} (\bibinfo {year}
  {2017})}\BibitemShut {NoStop}%
\bibitem [{\citenamefont {Yunusova}\ \emph {et~al.}(2019)\citenamefont
  {Yunusova}, \citenamefont {Konstantinov}, \citenamefont {Bouchiat},\ and\
  \citenamefont {Chepelianskii}}]{Yunusova2019}%
  \BibitemOpen
  \bibfield  {author} {\bibinfo {author} {\bibfnamefont {K.~M.}\ \bibnamefont
  {Yunusova}}, \bibinfo {author} {\bibfnamefont {D.}~\bibnamefont
  {Konstantinov}}, \bibinfo {author} {\bibfnamefont {H.}~\bibnamefont
  {Bouchiat}},\ and\ \bibinfo {author} {\bibfnamefont {A.~D.}\ \bibnamefont
  {Chepelianskii}},\ }\bibfield  {title} {\bibinfo {title} {Coupling between
  {{Rydberg States}} and {{Landau Levels}} of {{Electrons Trapped}} on {{Liquid
  Helium}}},\ }\href {https://doi.org/10.1103/PhysRevLett.122.176802}
  {\bibfield  {journal} {\bibinfo  {journal} {Phys. Rev. Lett.}\ }\textbf
  {\bibinfo {volume} {122}},\ \bibinfo {pages} {176802} (\bibinfo {year}
  {2019})}\BibitemShut {NoStop}%
\bibitem [{\citenamefont {Stoneham}(2001)}]{Stoneham2001}%
  \BibitemOpen
  \bibfield  {author} {\bibinfo {author} {\bibfnamefont {A.~M.}\ \bibnamefont
  {Stoneham}},\ }\href@noop {} {\emph {\bibinfo {title} {Theory of {{Defects}}
  in {{Solids}}}}}\ (\bibinfo  {publisher} {{Oxford University Press,
  Oxford}},\ \bibinfo {year} {2001})\BibitemShut {NoStop}%
\bibitem [{\citenamefont {Awschalom}\ \emph {et~al.}(2018)\citenamefont
  {Awschalom}, \citenamefont {Hanson}, \citenamefont {Wrachtrup},\ and\
  \citenamefont {Zhou}}]{Awschalom2018}%
  \BibitemOpen
  \bibfield  {author} {\bibinfo {author} {\bibfnamefont {D.~D.}\ \bibnamefont
  {Awschalom}}, \bibinfo {author} {\bibfnamefont {R.}~\bibnamefont {Hanson}},
  \bibinfo {author} {\bibfnamefont {J.}~\bibnamefont {Wrachtrup}},\ and\
  \bibinfo {author} {\bibfnamefont {B.~B.}\ \bibnamefont {Zhou}},\ }\bibfield
  {title} {\bibinfo {title} {Quantum technologies with optically interfaced
  solid-state spins},\ }\href {https://doi.org/10.1038/s41566-018-0232-2}
  {\bibfield  {journal} {\bibinfo  {journal} {Nat. Photonics}\ }\textbf
  {\bibinfo {volume} {12}},\ \bibinfo {pages} {516} (\bibinfo {year}
  {2018})}\BibitemShut {NoStop}%
\bibitem [{\citenamefont {Platzman}\ and\ \citenamefont
  {Dykman}(1999)}]{Platzman1999}%
  \BibitemOpen
  \bibfield  {author} {\bibinfo {author} {\bibfnamefont {P.~M.}\ \bibnamefont
  {Platzman}}\ and\ \bibinfo {author} {\bibfnamefont {M.~I.}\ \bibnamefont
  {Dykman}},\ }\bibfield  {title} {\bibinfo {title} {Quantum {{Computing}} with
  {{Electrons Floating}} on {{Liquid Helium}}},\ }\href@noop {} {\bibfield
  {journal} {\bibinfo  {journal} {Science}\ }\textbf {\bibinfo {volume}
  {284}},\ \bibinfo {pages} {1967} (\bibinfo {year} {1999})}\BibitemShut
  {NoStop}%
\bibitem [{\citenamefont {Lyon}(2006)}]{Lyon2006}%
  \BibitemOpen
  \bibfield  {author} {\bibinfo {author} {\bibfnamefont {S.~A.}\ \bibnamefont
  {Lyon}},\ }\bibfield  {title} {\bibinfo {title} {Spin-{{Based Quantum
  Computing Using Electrons}} on {{Liquid Helium}}},\ }\href@noop {} {\bibfield
   {journal} {\bibinfo  {journal} {Phys Rev A}\ }\textbf {\bibinfo {volume}
  {74}},\ \bibinfo {pages} {052338} (\bibinfo {year} {2006})}\BibitemShut
  {NoStop}%
\bibitem [{\citenamefont {Schuster}\ \emph {et~al.}(2010)\citenamefont
  {Schuster}, \citenamefont {Fragner}, \citenamefont {Dykman}, \citenamefont
  {Lyon},\ and\ \citenamefont {Schoelkopf}}]{Schuster2010}%
  \BibitemOpen
  \bibfield  {author} {\bibinfo {author} {\bibfnamefont {D.~I.}\ \bibnamefont
  {Schuster}}, \bibinfo {author} {\bibfnamefont {A.}~\bibnamefont {Fragner}},
  \bibinfo {author} {\bibfnamefont {M.~I.}\ \bibnamefont {Dykman}}, \bibinfo
  {author} {\bibfnamefont {S.~A.}\ \bibnamefont {Lyon}},\ and\ \bibinfo
  {author} {\bibfnamefont {R.~J.}\ \bibnamefont {Schoelkopf}},\ }\bibfield
  {title} {\bibinfo {title} {Proposal for {{Manipulating}} and {{Detecting
  Spin}} and {{Orbital States}} of {{Trapped Electrons}} on {{Helium Using
  Cavity Quantum Electrodynamics}}},\ }\href@noop {} {\bibfield  {journal}
  {\bibinfo  {journal} {Phys Rev Lett}\ }\textbf {\bibinfo {volume} {105}},\
  \bibinfo {pages} {040503} (\bibinfo {year} {2010})}\BibitemShut {NoStop}%
\bibitem [{\citenamefont {Yang}\ \emph {et~al.}(2016)\citenamefont {Yang},
  \citenamefont {Fragner}, \citenamefont {Koolstra}, \citenamefont {Ocola},
  \citenamefont {Czaplewski}, \citenamefont {Schoelkopf},\ and\ \citenamefont
  {Schuster}}]{Yang2016a}%
  \BibitemOpen
  \bibfield  {author} {\bibinfo {author} {\bibfnamefont {G.}~\bibnamefont
  {Yang}}, \bibinfo {author} {\bibfnamefont {A.}~\bibnamefont {Fragner}},
  \bibinfo {author} {\bibfnamefont {G.}~\bibnamefont {Koolstra}}, \bibinfo
  {author} {\bibfnamefont {L.}~\bibnamefont {Ocola}}, \bibinfo {author}
  {\bibfnamefont {D.~A.}\ \bibnamefont {Czaplewski}}, \bibinfo {author}
  {\bibfnamefont {R.~J.}\ \bibnamefont {Schoelkopf}},\ and\ \bibinfo {author}
  {\bibfnamefont {D.~I.}\ \bibnamefont {Schuster}},\ }\bibfield  {title}
  {\bibinfo {title} {Coupling an {{Ensemble}} of {{Electrons}} on {{Superfluid
  Helium}} to a {{Superconducting Circuit}}},\ }\href
  {https://doi.org/10.1103/PhysRevX.6.011031} {\bibfield  {journal} {\bibinfo
  {journal} {Phys. Rev. X}\ }\textbf {\bibinfo {volume} {6}},\ \bibinfo {pages}
  {011031} (\bibinfo {year} {2016})}\BibitemShut {NoStop}%
\bibitem [{\citenamefont {Byeon}\ \emph {et~al.}(2020)\citenamefont {Byeon},
  \citenamefont {Nasyedkin}, \citenamefont {Lane}, \citenamefont {Beysengulov},
  \citenamefont {Zhang}, \citenamefont {Loloee},\ and\ \citenamefont
  {Pollanen}}]{Byeon2020}%
  \BibitemOpen
  \bibfield  {author} {\bibinfo {author} {\bibfnamefont {H.}~\bibnamefont
  {Byeon}}, \bibinfo {author} {\bibfnamefont {K.}~\bibnamefont {Nasyedkin}},
  \bibinfo {author} {\bibfnamefont {J.~R.}\ \bibnamefont {Lane}}, \bibinfo
  {author} {\bibfnamefont {N.~R.}\ \bibnamefont {Beysengulov}}, \bibinfo
  {author} {\bibfnamefont {L.}~\bibnamefont {Zhang}}, \bibinfo {author}
  {\bibfnamefont {R.}~\bibnamefont {Loloee}},\ and\ \bibinfo {author}
  {\bibfnamefont {J.}~\bibnamefont {Pollanen}},\ }\bibfield  {title} {\bibinfo
  {title} {Piezoacoustics for flying electron qubits on helium},\ }\href@noop
  {} {\bibfield  {journal} {\bibinfo  {journal} {ArXiv200802330 Cond-Mat
  Physicsquant-Ph}\ } (\bibinfo {year} {2020})},\ \bibinfo {note} {comment:
  Main manuscript: 12 pages, 3 figures; Supplemental Information: 9 pages, 3
  figures, 1 table},\ \Eprint {https://arxiv.org/abs/2008.02330}
  {arXiv:2008.02330 [cond-mat, physics:quant-ph]} \BibitemShut {NoStop}%
\bibitem [{\citenamefont {Grimes}\ and\ \citenamefont
  {Adams}(1979)}]{Grimes1979}%
  \BibitemOpen
  \bibfield  {author} {\bibinfo {author} {\bibfnamefont {C.~C.}\ \bibnamefont
  {Grimes}}\ and\ \bibinfo {author} {\bibfnamefont {G.}~\bibnamefont {Adams}},\
  }\bibfield  {title} {\bibinfo {title} {Evidence {{For A Liquid}}-to-{{Crystal
  Phase}}-{{Transition In A Classical}}, 2-{{Dimensional Sheet}} of
  {{Electrons}}},\ }\href@noop {} {\bibfield  {journal} {\bibinfo  {journal}
  {Phys Rev Lett}\ }\textbf {\bibinfo {volume} {42}},\ \bibinfo {pages} {795}
  (\bibinfo {year} {1979})}\BibitemShut {NoStop}%
\bibitem [{\citenamefont {Fisher}\ \emph {et~al.}(1979)\citenamefont {Fisher},
  \citenamefont {Halperin},\ and\ \citenamefont {Platzman}}]{Fisher1979}%
  \BibitemOpen
  \bibfield  {author} {\bibinfo {author} {\bibfnamefont {D.~S.}\ \bibnamefont
  {Fisher}}, \bibinfo {author} {\bibfnamefont {B.~I.}\ \bibnamefont
  {Halperin}},\ and\ \bibinfo {author} {\bibfnamefont {P.~M.}\ \bibnamefont
  {Platzman}},\ }\bibfield  {title} {\bibinfo {title} {Phonon-{{Ripplon
  Coupling}} and the 2-{{Dimensional Electron Solid On A Liquid}}-{{Helium
  Surface}}},\ }\href@noop {} {\bibfield  {journal} {\bibinfo  {journal} {Phys
  Rev Lett}\ }\textbf {\bibinfo {volume} {42}},\ \bibinfo {pages} {798}
  (\bibinfo {year} {1979})}\BibitemShut {NoStop}%
\bibitem [{\citenamefont {Dykman}\ and\ \citenamefont
  {Khazan}(1979)}]{Dykman1979a}%
  \BibitemOpen
  \bibfield  {author} {\bibinfo {author} {\bibfnamefont {M.~I.}\ \bibnamefont
  {Dykman}}\ and\ \bibinfo {author} {\bibfnamefont {L.~S.}\ \bibnamefont
  {Khazan}},\ }\bibfield  {title} {\bibinfo {title} {Effect of the
  {{Interaction}} between {{Nondegenerate Electrons Localized}} in a {{Thin
  Surface Layer}} on the {{Cyclotron Resonance}} and on the
  {{Magnetoconductance}}},\ }\href@noop {} {\bibfield  {journal} {\bibinfo
  {journal} {JETP}\ }\textbf {\bibinfo {volume} {50}},\ \bibinfo {pages} {747}
  (\bibinfo {year} {1979})}\BibitemShut {NoStop}%
\bibitem [{\citenamefont {Edelman}(1980)}]{Edelman1980}%
  \BibitemOpen
  \bibfield  {author} {\bibinfo {author} {\bibfnamefont {V.~S.}\ \bibnamefont
  {Edelman}},\ }\bibfield  {title} {\bibinfo {title} {Levitating
  {{Electrons}}},\ }\href@noop {} {\bibfield  {journal} {\bibinfo  {journal}
  {Sov Phys Usp}\ }\textbf {\bibinfo {volume} {23}},\ \bibinfo {pages} {227}
  (\bibinfo {year} {1980})}\BibitemShut {NoStop}%
\bibitem [{\citenamefont {Wilen}\ and\ \citenamefont
  {Giannetta}(1988)}]{Wilen1988}%
  \BibitemOpen
  \bibfield  {author} {\bibinfo {author} {\bibfnamefont {L.}~\bibnamefont
  {Wilen}}\ and\ \bibinfo {author} {\bibfnamefont {R.}~\bibnamefont
  {Giannetta}},\ }\bibfield  {title} {\bibinfo {title} {Cyclotron-{{Resonance}}
  of the 2d {{Electron Crystal}}},\ }\href
  {https://doi.org/10.1016/0039-6028(88)90658-9} {\bibfield  {journal}
  {\bibinfo  {journal} {Surf Sci}\ }\textbf {\bibinfo {volume} {196}},\
  \bibinfo {pages} {24} (\bibinfo {year} {1988})}\BibitemShut {NoStop}%
\bibitem [{\citenamefont {Dykman}\ \emph {et~al.}(1993)\citenamefont {Dykman},
  \citenamefont {Lea}, \citenamefont {Fozooni},\ and\ \citenamefont
  {Frost}}]{Dykman1993b}%
  \BibitemOpen
  \bibfield  {author} {\bibinfo {author} {\bibfnamefont {M.~I.}\ \bibnamefont
  {Dykman}}, \bibinfo {author} {\bibfnamefont {M.~J.}\ \bibnamefont {Lea}},
  \bibinfo {author} {\bibfnamefont {P.}~\bibnamefont {Fozooni}},\ and\ \bibinfo
  {author} {\bibfnamefont {J.}~\bibnamefont {Frost}},\ }\bibfield  {title}
  {\bibinfo {title} {Magnetoresistance {{In 2D Electrons On Liquid}}-{{Helium}}
  - {{Many}}-{{Electron Versus Single}}-{{Electron Kinetics}}},\ }\href@noop {}
  {\bibfield  {journal} {\bibinfo  {journal} {Phys Rev Lett}\ }\textbf
  {\bibinfo {volume} {70}},\ \bibinfo {pages} {3975} (\bibinfo {year}
  {1993})}\BibitemShut {NoStop}%
\bibitem [{\citenamefont {Kristensen}\ \emph {et~al.}(1996)\citenamefont
  {Kristensen}, \citenamefont {Djerfi}, \citenamefont {Fozooni}, \citenamefont
  {Lea}, \citenamefont {Richardson}, \citenamefont {{Santrich-Badal}},
  \citenamefont {Blackburn},\ and\ \citenamefont {{van der
  Heijden}}}]{Kristensen1996}%
  \BibitemOpen
  \bibfield  {author} {\bibinfo {author} {\bibfnamefont {A.}~\bibnamefont
  {Kristensen}}, \bibinfo {author} {\bibfnamefont {K.}~\bibnamefont {Djerfi}},
  \bibinfo {author} {\bibfnamefont {P.}~\bibnamefont {Fozooni}}, \bibinfo
  {author} {\bibfnamefont {M.~J.}\ \bibnamefont {Lea}}, \bibinfo {author}
  {\bibfnamefont {P.~J.}\ \bibnamefont {Richardson}}, \bibinfo {author}
  {\bibfnamefont {A.}~\bibnamefont {{Santrich-Badal}}}, \bibinfo {author}
  {\bibfnamefont {A.}~\bibnamefont {Blackburn}},\ and\ \bibinfo {author}
  {\bibfnamefont {R.~W.}\ \bibnamefont {{van der Heijden}}},\ }\bibfield
  {title} {\bibinfo {title} {Hall-{{Velocity Limited Magnetoconductivity}} in a
  {{Classical Two}}-{{Dimensional Wigner Crystal}}},\ }\href@noop {} {\bibfield
   {journal} {\bibinfo  {journal} {Phys Rev Lett}\ }\textbf {\bibinfo {volume}
  {77}},\ \bibinfo {pages} {1350} (\bibinfo {year} {1996})}\BibitemShut
  {NoStop}%
\bibitem [{\citenamefont {Konstantinov}\ \emph {et~al.}(2013)\citenamefont
  {Konstantinov}, \citenamefont {Monarkha},\ and\ \citenamefont
  {Kono}}]{Konstantinov2013}%
  \BibitemOpen
  \bibfield  {author} {\bibinfo {author} {\bibfnamefont {D.}~\bibnamefont
  {Konstantinov}}, \bibinfo {author} {\bibfnamefont {Y.}~\bibnamefont
  {Monarkha}},\ and\ \bibinfo {author} {\bibfnamefont {K.}~\bibnamefont
  {Kono}},\ }\bibfield  {title} {\bibinfo {title} {Effect of {{Coulomb
  Interaction}} on {{Microwave}}-{{Induced Magnetoconductivity Oscillations}}
  of {{Surface Electrons}} on {{Liquid Helium}}},\ }\href
  {https://doi.org/10.1103/PhysRevLett.111.266802} {\bibfield  {journal}
  {\bibinfo  {journal} {Phys Rev Lett}\ }\textbf {\bibinfo {volume} {111}},\
  \bibinfo {pages} {266802} (\bibinfo {year} {2013})}\BibitemShut {NoStop}%
\bibitem [{\citenamefont {Chepelianskii}\ \emph {et~al.}(2015)\citenamefont
  {Chepelianskii}, \citenamefont {Watanabe}, \citenamefont {Nasyedkin},
  \citenamefont {Kono},\ and\ \citenamefont
  {Konstantinov}}]{Chepelianskii2015}%
  \BibitemOpen
  \bibfield  {author} {\bibinfo {author} {\bibfnamefont {A.~D.}\ \bibnamefont
  {Chepelianskii}}, \bibinfo {author} {\bibfnamefont {M.}~\bibnamefont
  {Watanabe}}, \bibinfo {author} {\bibfnamefont {K.}~\bibnamefont {Nasyedkin}},
  \bibinfo {author} {\bibfnamefont {K.}~\bibnamefont {Kono}},\ and\ \bibinfo
  {author} {\bibfnamefont {D.}~\bibnamefont {Konstantinov}},\ }\bibfield
  {title} {\bibinfo {title} {An {{Incompressible State}} of a
  {{Photo}}-{{Excited Electron Gas}}},\ }\href@noop {} {\bibfield  {journal}
  {\bibinfo  {journal} {Nat Commun}\ }\textbf {\bibinfo {volume} {6}},\
  \bibinfo {pages} {7210} (\bibinfo {year} {2015})}\BibitemShut {NoStop}%
\bibitem [{\citenamefont {Rees}\ \emph {et~al.}(2016)\citenamefont {Rees},
  \citenamefont {Beysengulov}, \citenamefont {Lin},\ and\ \citenamefont
  {Kono}}]{Rees2016}%
  \BibitemOpen
  \bibfield  {author} {\bibinfo {author} {\bibfnamefont {D.~G.}\ \bibnamefont
  {Rees}}, \bibinfo {author} {\bibfnamefont {N.~R.}\ \bibnamefont
  {Beysengulov}}, \bibinfo {author} {\bibfnamefont {J.-J.}\ \bibnamefont
  {Lin}},\ and\ \bibinfo {author} {\bibfnamefont {K.}~\bibnamefont {Kono}},\
  }\bibfield  {title} {\bibinfo {title} {Stick-{{Slip Motion}} of the {{Wigner
  Solid}} on {{Liquid Helium}}},\ }\href@noop {} {\bibfield  {journal}
  {\bibinfo  {journal} {Phys Rev Lett}\ }\textbf {\bibinfo {volume} {116}},\
  \bibinfo {pages} {206801} (\bibinfo {year} {2016})}\BibitemShut {NoStop}%
\bibitem [{\citenamefont {Barry}\ \emph {et~al.}(2020)\citenamefont {Barry},
  \citenamefont {Schloss}, \citenamefont {Bauch}, \citenamefont {Turner},
  \citenamefont {Hart}, \citenamefont {Pham},\ and\ \citenamefont
  {Walsworth}}]{Barry2020}%
  \BibitemOpen
  \bibfield  {author} {\bibinfo {author} {\bibfnamefont {J.~F.}\ \bibnamefont
  {Barry}}, \bibinfo {author} {\bibfnamefont {J.~M.}\ \bibnamefont {Schloss}},
  \bibinfo {author} {\bibfnamefont {E.}~\bibnamefont {Bauch}}, \bibinfo
  {author} {\bibfnamefont {M.~J.}\ \bibnamefont {Turner}}, \bibinfo {author}
  {\bibfnamefont {C.~A.}\ \bibnamefont {Hart}}, \bibinfo {author}
  {\bibfnamefont {L.~M.}\ \bibnamefont {Pham}},\ and\ \bibinfo {author}
  {\bibfnamefont {R.~L.}\ \bibnamefont {Walsworth}},\ }\bibfield  {title}
  {\bibinfo {title} {Sensitivity optimization for {{NV}}-diamond
  magnetometry},\ }\href {https://doi.org/10.1103/RevModPhys.92.015004}
  {\bibfield  {journal} {\bibinfo  {journal} {Rev. Mod. Phys.}\ }\textbf
  {\bibinfo {volume} {92}},\ \bibinfo {pages} {015004} (\bibinfo {year}
  {2020})}\BibitemShut {NoStop}%
\bibitem [{\citenamefont {Bhaskar}\ \emph {et~al.}(2020)\citenamefont
  {Bhaskar}, \citenamefont {Riedinger}, \citenamefont {Machielse},
  \citenamefont {Levonian}, \citenamefont {Nguyen}, \citenamefont {Knall},
  \citenamefont {Park}, \citenamefont {Englund}, \citenamefont {Lon{\v c}ar},
  \citenamefont {Sukachev},\ and\ \citenamefont {Lukin}}]{Bhaskar2020}%
  \BibitemOpen
  \bibfield  {author} {\bibinfo {author} {\bibfnamefont {M.~K.}\ \bibnamefont
  {Bhaskar}}, \bibinfo {author} {\bibfnamefont {R.}~\bibnamefont {Riedinger}},
  \bibinfo {author} {\bibfnamefont {B.}~\bibnamefont {Machielse}}, \bibinfo
  {author} {\bibfnamefont {D.~S.}\ \bibnamefont {Levonian}}, \bibinfo {author}
  {\bibfnamefont {C.~T.}\ \bibnamefont {Nguyen}}, \bibinfo {author}
  {\bibfnamefont {E.~N.}\ \bibnamefont {Knall}}, \bibinfo {author}
  {\bibfnamefont {H.}~\bibnamefont {Park}}, \bibinfo {author} {\bibfnamefont
  {D.}~\bibnamefont {Englund}}, \bibinfo {author} {\bibfnamefont
  {M.}~\bibnamefont {Lon{\v c}ar}}, \bibinfo {author} {\bibfnamefont {D.~D.}\
  \bibnamefont {Sukachev}},\ and\ \bibinfo {author} {\bibfnamefont {M.~D.}\
  \bibnamefont {Lukin}},\ }\bibfield  {title} {\bibinfo {title} {Experimental
  demonstration of memory-enhanced quantum communication},\ }\href
  {https://doi.org/10.1038/s41586-020-2103-5} {\bibfield  {journal} {\bibinfo
  {journal} {Nature}\ }\textbf {\bibinfo {volume} {580}},\ \bibinfo {pages}
  {60} (\bibinfo {year} {2020})}\BibitemShut {NoStop}%
\bibitem [{\citenamefont {Ando}\ \emph {et~al.}(1982)\citenamefont {Ando},
  \citenamefont {Fowler},\ and\ \citenamefont {Stern}}]{Ando1982}%
  \BibitemOpen
  \bibfield  {author} {\bibinfo {author} {\bibfnamefont {T.}~\bibnamefont
  {Ando}}, \bibinfo {author} {\bibfnamefont {A.~B.}\ \bibnamefont {Fowler}},\
  and\ \bibinfo {author} {\bibfnamefont {F.}~\bibnamefont {Stern}},\ }\bibfield
   {title} {\bibinfo {title} {Electronic-{{Properties Of Two}}-{{Dimensional
  Systems}}},\ }\href@noop {} {\bibfield  {journal} {\bibinfo  {journal} {Rev
  Mod Phys}\ }\textbf {\bibinfo {volume} {54}},\ \bibinfo {pages} {437}
  (\bibinfo {year} {1982})}\BibitemShut {NoStop}%
\bibitem [{\citenamefont {Zipfel}\ \emph
  {et~al.}(1976{\natexlab{a}})\citenamefont {Zipfel}, \citenamefont {Brown},\
  and\ \citenamefont {Grimes}}]{Zipfel1976}%
  \BibitemOpen
  \bibfield  {author} {\bibinfo {author} {\bibfnamefont {C.~L.}\ \bibnamefont
  {Zipfel}}, \bibinfo {author} {\bibfnamefont {T.~R.}\ \bibnamefont {Brown}},\
  and\ \bibinfo {author} {\bibfnamefont {C.~C.}\ \bibnamefont {Grimes}},\
  }\bibfield  {title} {\bibinfo {title} {Spectroscopic studies of electron
  surface states on liquid helium},\ }\href
  {https://doi.org/10.1016/0039-6028(76)90151-5} {\bibfield  {journal}
  {\bibinfo  {journal} {Surf. Sci.}\ }\textbf {\bibinfo {volume} {58}},\
  \bibinfo {pages} {283} (\bibinfo {year} {1976}{\natexlab{a}})}\BibitemShut
  {NoStop}%
\bibitem [{\citenamefont {Zipfel}\ \emph
  {et~al.}(1976{\natexlab{b}})\citenamefont {Zipfel}, \citenamefont {Brown},\
  and\ \citenamefont {Grimes}}]{Zipfel1976a}%
  \BibitemOpen
  \bibfield  {author} {\bibinfo {author} {\bibfnamefont {C.~L.}\ \bibnamefont
  {Zipfel}}, \bibinfo {author} {\bibfnamefont {T.~R.}\ \bibnamefont {Brown}},\
  and\ \bibinfo {author} {\bibfnamefont {C.~C.}\ \bibnamefont {Grimes}},\
  }\bibfield  {title} {\bibinfo {title} {Measurement of the {{Velocity
  Autocorrelation Time}} in a {{Two}}-{{Dimensional Electron Liquid}}},\ }\href
  {https://doi.org/10.1103/PhysRevLett.37.1760} {\bibfield  {journal} {\bibinfo
   {journal} {Phys. Rev. Lett.}\ }\textbf {\bibinfo {volume} {37}},\ \bibinfo
  {pages} {1760} (\bibinfo {year} {1976}{\natexlab{b}})}\BibitemShut {NoStop}%
\bibitem [{\citenamefont {Pekar}(1950)}]{Pekar1950}%
  \BibitemOpen
  \bibfield  {author} {\bibinfo {author} {\bibfnamefont {S.~I.}\ \bibnamefont
  {Pekar}},\ }\bibfield  {title} {\bibinfo {title} {Theory of {{Color
  Centers}}},\ }\href@noop {} {\bibfield  {journal} {\bibinfo  {journal} {Zh
  Eksper Teor Fiz}\ }\textbf {\bibinfo {volume} {20}},\ \bibinfo {pages} {510}
  (\bibinfo {year} {1950})}\BibitemShut {NoStop}%
\bibitem [{\citenamefont {Huang}\ and\ \citenamefont {Rhys}(1950)}]{Huang1950}%
  \BibitemOpen
  \bibfield  {author} {\bibinfo {author} {\bibfnamefont {K.}~\bibnamefont
  {Huang}}\ and\ \bibinfo {author} {\bibfnamefont {A.}~\bibnamefont {Rhys}},\
  }\bibfield  {title} {\bibinfo {title} {Theory of light absorption and
  non-radiative transitions in {{F}}-centres},\ }\href
  {https://doi.org/10.1098/rspa.1950.0184} {\bibfield  {journal} {\bibinfo
  {journal} {Proc. Roy. Soc. A}\ }\textbf {\bibinfo {volume} {204}},\ \bibinfo
  {pages} {406} (\bibinfo {year} {1950})}\BibitemShut {NoStop}%
\bibitem [{\citenamefont {Lambert}\ and\ \citenamefont
  {Richards}(1981)}]{Lambert1981}%
  \BibitemOpen
  \bibfield  {author} {\bibinfo {author} {\bibfnamefont {D.~K.}\ \bibnamefont
  {Lambert}}\ and\ \bibinfo {author} {\bibfnamefont {P.~L.}\ \bibnamefont
  {Richards}},\ }\bibfield  {title} {\bibinfo {title} {Far-{{Infrared}} and
  {{Capacitance Measurements}} of {{Electrons}} on {{Liquid Helium}}},\
  }\href@noop {} {\bibfield  {journal} {\bibinfo  {journal} {Phys Rev B}\
  }\textbf {\bibinfo {volume} {23}},\ \bibinfo {pages} {3282} (\bibinfo {year}
  {1981})}\BibitemShut {NoStop}%
\bibitem [{Note1()}]{Note1}%
  \BibitemOpen
  \bibinfo {note} {The Supplemental Material provides the details of the
  calculation and the relation to the SI units}\BibitemShut {NoStop}%
\bibitem [{Note2()}]{Note2}%
  \BibitemOpen
  \bibinfo {note} {We are using CGS units. To switch to the SI units, one
  should set in the expressions we give $c=1$, replace $e^2\to e^2/4\pi
  \epsilon _0$ and, in the main text, in Eq.~(\ref {eq:central_peak}) for
  $w(t)$ and Eq.~(\ref {eq:Gaussian_width}), replace $n_s^{3/2} \to
  n_s^{3/2}/4\pi \epsilon _0$.}\BibitemShut {Stop}%
\bibitem [{\citenamefont {FangYen}\ \emph {et~al.}(1997)\citenamefont
  {FangYen}, \citenamefont {Dykman},\ and\ \citenamefont {Lea}}]{FangYen1997}%
  \BibitemOpen
  \bibfield  {author} {\bibinfo {author} {\bibfnamefont {C.}~\bibnamefont
  {FangYen}}, \bibinfo {author} {\bibfnamefont {M.~I.}\ \bibnamefont
  {Dykman}},\ and\ \bibinfo {author} {\bibfnamefont {M.~J.}\ \bibnamefont
  {Lea}},\ }\bibfield  {title} {\bibinfo {title} {Internal {{Forces}} in
  {{Nondegenerate Two}}-{{Dimensional Electron Systems}}},\ }\href@noop {}
  {\bibfield  {journal} {\bibinfo  {journal} {Phys Rev B}\ }\textbf {\bibinfo
  {volume} {55}},\ \bibinfo {pages} {16272} (\bibinfo {year}
  {1997})}\BibitemShut {NoStop}%
\bibitem [{\citenamefont {Moskovtsev}\ and\ \citenamefont
  {Dykman}(2019)}]{Moskovtsev2019}%
  \BibitemOpen
  \bibfield  {author} {\bibinfo {author} {\bibfnamefont {K.}~\bibnamefont
  {Moskovtsev}}\ and\ \bibinfo {author} {\bibfnamefont {M.~I.}\ \bibnamefont
  {Dykman}},\ }\bibfield  {title} {\bibinfo {title} {Self-{{Diffusion}} in a
  {{Spatially Modulated System}} of {{Electrons}} on {{Helium}}},\ }\href
  {https://doi.org/10.1007/s10909-019-02148-z} {\bibfield  {journal} {\bibinfo
  {journal} {J. Low Temp. Phys.}\ }\textbf {\bibinfo {volume} {195}},\ \bibinfo
  {pages} {266} (\bibinfo {year} {2019})}\BibitemShut {NoStop}%
\bibitem [{\citenamefont {Ulinich}\ and\ \citenamefont
  {Usov}(1979)}]{Ulinich1979}%
  \BibitemOpen
  \bibfield  {author} {\bibinfo {author} {\bibfnamefont {F.}~\bibnamefont
  {Ulinich}}\ and\ \bibinfo {author} {\bibfnamefont {N.}~\bibnamefont {Usov}},\
  }\bibfield  {title} {\bibinfo {title} {Phase {{Diagram}} of a
  {{Two}}-{{Dimensional Wigner Crystal}} in a {{Magnetic Field}}},\ }\href@noop
  {} {\bibfield  {journal} {\bibinfo  {journal} {JETP}\ }\textbf {\bibinfo
  {volume} {49}},\ \bibinfo {pages} {147} (\bibinfo {year} {1979})}\BibitemShut
  {NoStop}%
\end{thebibliography}

\end{document}